\documentclass[twocolumn,secnumarabic,amssymb, aps, prl]{revtex4-1}
\usepackage{graphicx}
\usepackage{amsmath}
\usepackage{amsfonts}
\usepackage{multirow}
\usepackage[colorlinks, linkcolor=blue, anchorcolor=blue,citecolor=blue]{hyperref}

\begin{document}
	\title{Non-Fermi-Liquid/Marginal-Fermi-Liquid Signatures Induced by Van Hove Singularity}%
	\author{Yi-Hui Xing$^{1,2}$}
	\author{Wu-Ming Liu$^{1,2,3}$}
	\email{wliu@iphy.ac.cn}
	\affiliation{$^{1}$Beijing National Laboratory for Condensed Matter Physics, Institute of Physics, Chinese Academy of Sciences, Beijing 100190, China}
	\affiliation{$^{2}$School of Physical Sciences, University of Chinese Academy of Sciences, Beijing 100190, China}
	\affiliation{$^{3}$Songshan Lake Materials Laboratory, Dongguan, Guangdong 523808, China}

	\begin{abstract}
	We theoretically study the two-dimensional metal that is coupled to critical magnons and features van Hove singularities on the Fermi surface. When there is only translationally invariant SYK-liked Yukawa interaction, van Hove points suppress the contribution from the part of the Fermi surface away from them, dominating and exhibiting non-Fermi-liquid behavior. When introducing disordered Yukawa coupling, it leads to a crossover from non-Fermi-liquid to marginal-Fermi-liquid, and the marginal-Fermi-liquid region exhibits the $T\ln (1/T)$ specific heat and temperature-linear resistivity of strange metal. By solving the gap equation, we provide the critical temperature for superconductor induced by van Hove singularities and point out the possible emergence of pair-density-wave superconductor. Our theory may become a new mechanism for understanding non-Fermi-liquid or marginal-Fermi-liquid phenomenons.
	\end{abstract}
	
	\maketitle
	
	Many strongly correlated systems can exhibit phenomenons deviating from the predictions of the Fermi-liquid theory of metals, namely, non-Fermi-liquid (NFL) or marginal-Fermi-liquid (MFL) behavior. The most prominent examples include iron-based superconductors \cite{reiss2020quenched, PhysRevB.86.245113}, heavy fermion materials \cite{chen2019heavy, seiro2018evolution}, and moir\'{e} system \cite{PhysRevB.101.205426, wang2022one, PhysRevLett.124.076801}, among others. A direct signature of NFL is the frequency $\omega$ and temperature $T$ dependent relationship of self-energy imaginary part $\mathrm{Im}\Sigma(\omega,T)$ of electrons behaves for $\omega>T$ like $\sim |\omega|^\alpha$, where $\alpha<1$. When $\alpha = 1$ \cite{PhysRevLett.63.1996, PhysRevLett.94.156401}, it corresponds to MFL, whereas for ordinary Fermi-liquid, $\alpha = 2$. Because NFL/MFL is defined by deviating from the Fermi-liquid, this has led to a diverse family of systems, and lacking universal experimental features akin to the Fermi liquid. The microscopic origins of the NFL/MFL phenomena have been a fascinating open issue in condensed matter physics in recent years.
	
	Using the critical Fermi surface \cite{CFM} can characterize the low-energy effective theory of NFL/MFL, where free fermions couple with critical bosons. However, most of the previous work has been based on Fermi surfaces away from half-filling \cite{science.abq6011, PhysRevB.103.235129, PhysRevB.106.115151, physRevB.80.165102} or hot spots \cite{PhysRevB.90.161106, PhysRevLett.130.083603}, i.e., without considering the influence of van Hove singularities (VHS). Considering that the Fermi surfaces of many materials \cite{polshyn2019large, PhysRevLett.124.076801, mousatov2020theory, xu2021magnetotransport, wang2020correlated} exhibiting NFL/MFL phenomena in experiments include van Hove points (VHPs) and dopping can modify the Fermi surface to introduce VHS, our work investigated the impact of VHS on NFL/MFL behavior.
	
	VHPs are points of divergent density of states. Due to the singularities \cite{PhysRevLett.112.070403} of the electron-electron scattering close to VHPs, the Fermi-liquid picture is violated, leading to the emergence of various phases such as superconductors (SC) and density waves. The Fermi levels of many materials \cite{PhysRevLett.125.166401, kang2022twofold, wu2021chern, luo2023unique} exhibit VHS, it has been theoretically employed to explain phenomena such as high-temperature superconductivity \cite{PhysRevB.93.094525, PhysRevLett.89.076401} and pair-density-wave \cite{PhysRevLett.130.126001, PhysRevLett.131.026601}. VHS also alters the signatures of critical bosons, modifying the low-energy and temperature properties, which can, in turn, lead to NFL/MFL behavior.

	In this Letter, we study a system with electron-spin exchange interactions, where the Fermi surface exhibits VHS. We extending it to a SYK-liked \cite{PhysRevLett.70.3339, PhysRevLett.124.017002, PhysRevX.8.031024, PhysRevLett.130.026001} flavor-random Yukawa interactions, it describes $N$ flavors of fermion couple with $N'$ critical bosons. We show that when the interaction is translationally invariant, the system will exhibit NFL behavior, and VHS will suppress the behavior of the regular Fermi surface (i.e., the Fermi surface away from VHS) and play a dominant role in NFL behavior. When breaking translational invariance, i.e., introducing impurities, the system will exhibit MFL behavior. The temperature $T$-linear resistivity and $T\ln(1/T)$ electronic specific heat implies the emergence of a strange metal \cite{mousatov2020theory, science.abq6011, PhysRevX.5.041025, PhysRevLett.124.076801, chen2019incoherent, RevModPhys.94.035004}. When both types of interactions are considered, crossovers will appear and depend on the energy scale or behavior of $N/N'$. For completeness, we numerically identify the finite-temperature phase boundaries of the boson modes by setting the effective tuning parameter to zero \cite{phi4}. Using the gap equation, we identify the critical temperature for the SC. We observed completely different tuning parameter $\Delta$ dependent behaviors for the critical temperatures of the two interactions. For real systems, the coexistence of disorder interaction and translational invariant one, strange metal phase could also appear above the SC phase. We argue that the pair-density-wave (PDW) \cite{PhysRevLett.130.126001, zhou2022chern, hamidian2016detection, PhysRevLett.114.197001} order will appear to compete with conventional SC when the Fermi surface contains more than one VHPs.  
	
\begin{figure}
		\centering
		\includegraphics[width=\columnwidth]{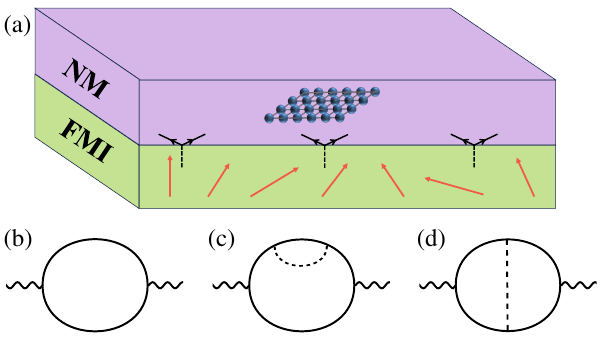}
		\caption{(a) An illustration of the bilayer system. The itinerant electrons in the two-dimensional NM interact with the spins (red arrows) in the FMI via exchange coupling (black Feynman diagram). The NM at the interface is modeled by lattice system (blue array). (b)-(d) The contributing conductivity diagrams. Lines represent fermionic propagators, wave lines represent current operators, dashed lines represent magnon propagators after damped.}
		\label{fig1}
	\end{figure}

	We consider a two-dimensional normal metal (NM) couple to a ferromagnetic insulator (TMI). The system is shown in Fig. \ref{fig1}(a) and the interface is described by the Hamiltonian
	\begin{equation}
		H=-\sum_{i,j,\sigma}t_{ij}c^\dagger_{i,\sigma}c_{j,\sigma}+\sum_{k}\omega_k a^\dagger_{k}a_{k}+H_{int},
		\label{ham}
	\end{equation}
	where $c_{i,\sigma}$ describes annihilating an electron at lattice site $i$ with spin $\sigma$, it constitutes the hopping term with energy $t_{ij}$. It is worth noting that we do not constrain the types of the lattice and the range of the hopping. The TMI is modeled by a nearest-neighbor ferromagnetic exchange interaction of strength $J$ and a tunable easy-axis anisotropy of strength $K$ \cite{PhysRevB.98.144411,PhysRevResearch.4.L032025} that can be tuned by applied mechanical strain.	The Hamiltonian of TMI is $H_{FI}=-\bar{J}\sum_{<ij>}\vec{S}_i\vec{S}_j-K\sum_iS_i^{z2}$ with local spin operator $S_i$. Performing the Holstein-Primakoff transformation in term of the magnon operator $a_k$, we get the magnon spectrum $\omega_k=S\bar{J}a^2k^2/2+2SK$ with the lattice constant $a$ of TMI and the spin $S$. For convenience, we set the magnon energy spectrum to be $\omega_k=k^2+\Delta$ (by transformation $a_k\rightarrow \sqrt{2/S\bar{J}a^2}a_k$), where $\Delta\equiv k_z^2+4K/\bar{J}a^2$. Note that we are considering a two-dimensional interface of a three-dimensional system, so the remaining momentum $k_z$ in the z-direction forms the tuning parameter $\Delta$. We neglect high order magnon-magnon interactions above and the subsequent magnon-electron interactions as they do not play a crucial role in our calculations.
	
	Electrons in the normal mental interact with spins in the ferromagnetic insulator at the interface via exchange $H_{em}=-\frac{J}{\sqrt{2S}}\sum_{i,\alpha,\beta}c_{i\alpha}^\dagger\vec{\sigma}^{\alpha\beta}c_{i\beta}\vec{S}_i$, where $\vec{\sigma}$ is vector of Pauli matrices. Applying the Holstein-Primakoff transformation and retained to the lowest order, yields $H_{em}=-J\sum_{i}(c_{i\uparrow}^\dagger c_{i\downarrow}a_i+\mathrm{h.c.})$. This interaction can be extended to the situation of quenched disorder after considering the influence of impurities, i.e. the coupling $J$ depends on the lattice site $i$. Experimentally, moving the Fermi surface through doping directly introduces impurities. The identical energy spectrum for electrons with different spin $\sigma$ \cite{ElEnSp} allows for an extension to the coupling between large-$N$ flavor of electrons. Similarly, we also consider the large-$N'$ flavor of magnons. We contemplate the following two types of coupling between electrons and magnons: 
	\begin{equation}
		\begin{aligned}
  			H_{int}=&-\frac{1}{\sqrt{NN'}}\sum_{mnl}J_{mnl}\sum_i c^\dagger_{i,m} c_{i,n} a_{i,l}\\
  			&-\frac{1}{\sqrt{NN'}}\sum_{mnl}\sum_i J'_{i,mnl} c^\dagger_{i,m} c_{i,n} a_{i,l}+h.c.,
  		\label{hint}
  		\end{aligned}
    \end{equation}	
	which includes translationally invariant (i.e. space independent) flavor random SYK-liked Yukawa coupling  and spatially dependent one, respectively. Where the space independent (dependent) random Yukawa couplings are chosen from a Gaussian unitary ensemble with zero average and variance $\overline{J_{mnl}J_{m'n'l'}}=|J|^2\delta_{mm'}\delta_{nn'}\delta_{ll'}$ ($\overline{J'_{i,mnl}J'_{i',m'n'l'}}=|J'|^2\delta_{ii'}\delta_{mm'}\delta_{nn'}\delta_{ll'}$), and $J_{mnl}=J^*_{nml}$ ($J'_{mnl}=J'^*_{nml}$). We assume that each sample flows toward a universal low-energy theory.
	
	The electrons and spin couplings lead to the phase transition from $U(1)$ phase to Bose-Einstein condensation (BEC) phase. The critical point of the phase transition at zero temperature is $D_0(0,Q)^{-1}=\Pi(0,Q)$, where $D_0(0,Q)$ is the bare propagator of magnons and $\Pi(0,Q)$ is the self-energy of magnons introduced by the interactions. Different hopping terms and crystal lattices in Eq. (\ref{ham}) can lead to rich band structures. We primarily consider the influence of scattering near VHPs, where the electron's dispersion relation exhibits either a maximum or a saddle-point on the Fermi surface, because VHS can lead to anomalous low-energy behavior. It can be anticipated that when filled around the Fermi surface with VHS, the physical properties are primarily determined by VHPs and are not significantly influenced by the overall shape of the Fermi surface \cite{EP}. When Taylor expanding the electron's energy spectrum at the VHPs, the first-order momentum term vanishes, and the second-order term appears. The saddle-point nature tells us that it can always be transformed into $\varepsilon_k=k_x^2+a k_y^2$ with $a<0$ through a coordinate transformation \cite{EnEl, PhysRevX.8.041041, PhysRevB.92.035132}, Where $k_x$ and $k_y$ are two orthogonal momentum directions after the coordinate transformation and the coefficient in front of $k_x^2$ is set to 1 through a reparameterization of the operator $c_k$. We only calculated the nematic order, which the order parameter momentum condenses at $Q=0$ and the critical point is $\Delta_c=\Pi(0,0)$, and subsequently demonstrate that our results also apply to density-wave orders (i.e. $Q\ne 0$). The divergence of the density of state at the VHP allows us to focus only on a small patch near it, the nematic order compels us to consider interactions within the same patch only.
	 
	 When considering only the translationally invariant SYK-liked Yukawa interaction (i.e. $J'=0$), we obtain the self-energy of electrons ($\Sigma$) and magnons ($\Pi$) at zero temperature and low-frequency limit (detailed derivations are shown in Supplemental Material \cite{SM}):
	 \begin{equation}
	 	\begin{aligned}
	 		&\Pi(i\Omega_m,q)-\Pi(0,0) \sim |J|^2\frac{N}{N'}f(q_x,q_y,-a)|\Omega_m|,\\
	 		&\Sigma(i\omega_n,0)-\Sigma(0,0)\sim -i|J|\sqrt{\frac{N'}{N}}\mathrm{sgn}(\omega_n)|\omega_n|^{1/2},
	 	\end{aligned}
	 	\label{sic-sef}
	 \end{equation}
	where $f(q_x,q_y,-a)=(|1-\sqrt{-a}\frac{q_y}{|q_x|}|-|1+\sqrt{-a}\frac{q_y}{|q_x|}|)/8\pi\sqrt{-a}(q_x^2+a q_y^2)$. The quasiparticle decay exhibits the scaling of NFL behavior. Compared to the quasiparticle decay results $\mathrm{Im}\Sigma(i\Omega_n,0)\sim \mathrm{sgn}(\omega_n)|\omega_n|^{2/3}$ \cite{science.abq6011,PhysRevLett.130.083603,PhysRevB.103.235129,PhysRevB.106.115151,PhysRevB.90.161106} obtained by considering scattering between patches around the hot spots in circular Fermi surface, which are far from the VHPs and the electrons dispersion is $\varepsilon_k=\pm v_Fk_\perp+\kappa k_\parallel^2/2$, scattering of patch near the VHP suppresses it and takes a dominant role. So, it is expected that when considering scattering over the entire Fermi surface with VHS, the quasiparticle decay rate will exhibits $\sim |\omega_n|^{1/2}$ NFL scaling.
	
	When considering only the spatially dependent interaction (i.e. $J=0$), we obtain the self-energy of electrons and magnons at zero temperature and low-frequency limit (detailed derivations are shown in Supplemental Material \cite{SM}):
	 \begin{equation}
		 \begin{aligned}
	 		&\Pi(i\Omega_m)-\Pi(0) \sim -|J'|^2\frac{N}{N'}|\Omega_m|,\\
	 		&\Sigma(i\omega_n)-\Sigma(0)\sim -i|J'|^2\omega_n \ln\frac{N'}{N|\omega_n|}.
	 	 \end{aligned}
	 	 \label{sdc-sef}
	 \end{equation}
	This quasiparticle decay exhibits MFL behavior and the specific heat is $\sim T\ln(1/T)$ \cite{PhysRevB.48.7297} with temperature $T$. Then we use the Kubo formula to calculate the direct current (dc) conductivity at $\mathcal{O}(|J'|^2)$. The related current-current corrections are shown in Fig. \ref{fig1}(b)-(d). Retaining up to the lowest order in frequency, the system resistivity \cite{SM} (i.e., the reciprocal of dc conductivity) is
	\begin{equation}
		\mathrm{Re}[\frac{1}{\sigma(\Omega\gg T)}]=\frac{16\pi^4(-a)|\Omega|}{|J'|^2N\Lambda_U^2\ln\Lambda},
	\end{equation}
	where $\Lambda_U$ is the momentum UV cutoff and $\Lambda$ is the ratio of the momentum UV cutoff to the frequency. Only self-energy correction in Fig. \ref{fig1}(c) contributes in this case. We only considered forward scattering of the patch near same VHP, which results in one-loop contribution Fig. \ref{fig1}(b) being zero. Meanwhile, the dispersion relation of electrons at the VHPs only contains the momentum squared term, ensuring that the vertex correction in Fig. \ref{fig1}(d) is zero. At the limit of $\Omega\ll T$, this will result in a $T$-linear resistivity in the dc limit, with no residual constant term. The $T(\ln 1/T)$ specific heat and $T$-linear resistivity are consistent with the behavior of strange metals.
	
	We can also obtain the above results of quasiparticle decay through the method of scaling analysis. After considering the VHS, the effect Lagrangian of our theory (\ref{ham}) is
	\begin{equation}
		\mathcal{L}=\psi^\dagger_m[\partial_\tau-\partial_x^2-a\partial_y^2]\psi+(\partial\phi_l)^2+\mathcal{H}_{int}
		\label{eff}
	\end{equation}
	with $\mathcal{H}_{int1}=\sum_{mnl}J_{mnl}\psi^\dagger_m\psi_n\phi_l/\sqrt{NN'}$ and $\mathcal{H}_{int2}=\sum_{mnl}J'_{mnl}(\vec{x})\psi^\dagger_m\psi_n\phi_l/\sqrt{NN'}$. If we assume the scaling dimension of space to be $[x]=[y]=1$, and time to be $[\tau]=z$, respectively. Then the scaling dimensions for the Fermi and Bose fields are both $[\psi]=[\phi]=z/2$. When we only consider the spatially independent interaction, we obtained the scaling dimension of the variance $J$ is $2-z/2$. NFL/MFL behavior occurs near the critical point, where we expect the interaction to be marginal, so $z=4$. This leads to the quasi-particle decay $\sim \omega^{1/2}$ and consists with the analytical results in Eq. (\ref{sic-sef}). Similarly, when considering only spatially dependent interaction, the scaling dimension of $J'$ is $1-z/2$, leading to $z=2$ and resulting in quasi-particle damping $\sim \omega$. This differs slightly from our analytical result $\omega\ln(1/\omega)$ in Eq. (\ref{sdc-sef}), but since the dominant role is played by the preceding linear term, it is also considered consistent. Our analytical calculations only considered scattering between the same VHP (i.e. Ising order). If we consider density-wave order with finite wave vector $\vec{Q}$, we need to account for scattering between patches around different VHPs, introducing more fermion modes such as $\psi^\dagger_{m1}[\partial_\tau-\partial_x^2-a\partial_y^2]\psi_{m1}$ and considering the interactions, $\psi^\dagger_m\psi_{m1}\phi_l e^{i\vec{Q}\vec{x}}$, between them. However, the scaling dimensions of the fields and couplings remain unchanged, so the same dynamic critical exponent, i.e., the same quasi-particle decay, will be obtained. This has already been verified \cite{PhysRevLett.130.083603, PhysRevB.103.235129} in the case considering only scattering at the Fermi surface away from VHPs.
	
	\begin{figure}
		\centering
		\includegraphics[width=\linewidth]{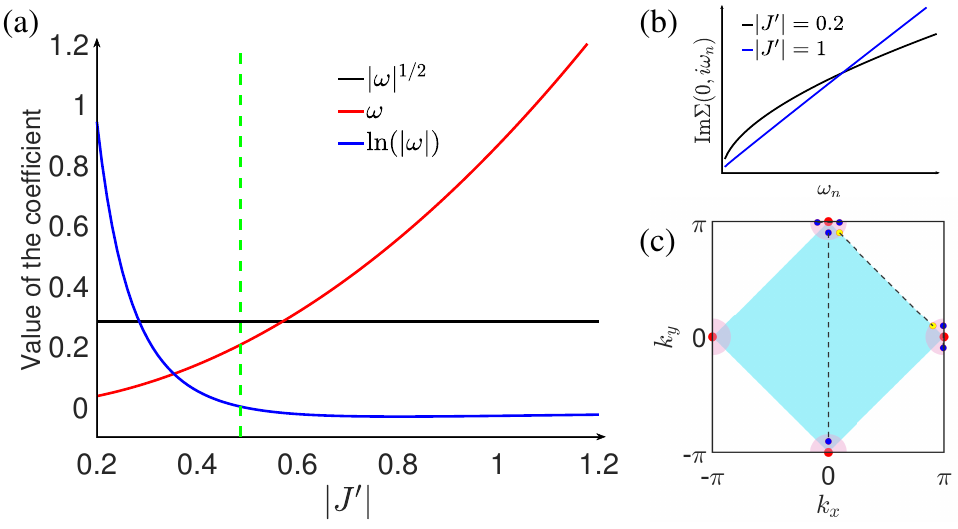}
		\caption{(a) The coefficients of different scaling as a function of the variance $|J'|$ of the disordered interaction. The black, red, and blue solid lines represent the coefficients of scaling of $|\omega|^{1/2}$, $\omega$, and $\ln (|\omega|)$, respectively. The green dashed line represents the critical $|J'|$ of two analytical limits. (b) The imaginary part of electron self-energy as a function of frequency $\omega_n$ when $|J'|=0.2$ and $1$. The other parameters in (a) and (b) are: $a=-1$, $\Lambda_{U}=300$, $\omega_n=21\pi$, $N=N'$ and $|J|=2$. (c) A schematic diagram of SC order and PDW order. The blue region represents the filled Fermi surface, red dots denote VHPs, and the surrounding red areas indicate the patches near them. The blue pairings represent conventional SC pairs formed near the same VHP, while the yellow pairings represent PDW pairs formed between different VHPs.}
		\label{fig2}
	\end{figure}
	
	In the presence of both types of interactions (i.e. $J\ne 0\ne J'$), the self-energy of magnons at zero temperature and low-frequency is the sum of the self-energy of magnons in (\ref{sic-sef}) and (\ref{sdc-sef}). The self-energy of electrons is shown in Supplemental Material \cite{SM}, different scalings are competing with each other. If we only consider the behavior of large $N$($N'$) and ignore the differences in energy scales. When $N\gg N'$, it exhibits quasiparticle decay rate $-\frac{|J'|^2\ln \Lambda}{2\pi^4\sqrt{-a}}[\mathrm{ArcTan}(\sqrt{-a})+	\mathrm{ArcTan}(\frac{1}{\sqrt{-a}})]\omega_n\ln\frac{e\pi^3(-a)N'\Lambda_U^2}{N(\ln \Lambda)^2|J'|^2|\omega_n|}$, and it is the scaling of MFL, it will also lead to the $T\ln (1/T)$ specific heat and the $T$-linear resistivity in the direct current limit; when $N\ll N'$, the scaling of NFL dominates and the quasiparticle decay rate $\sim |\omega_n|^{1/2}$. When  $N\approx N'$ and considering only the difference in energy scales, the situation becomes extremely complicated due to the appearance of logarithmic terms. As shown in Fig. \ref{fig2}(a) and \ref{fig2}(b), when $|J'|\ll |J|$, it exhibits the scaling of NFL. As $|J'|$ increases, there is a crossover to the scaling of MFL. Despite the appearance of the scaling with $\ln |\omega|$ in the analytical expression (appears on the right side of the green dashed line in Fig. \ref{fig2}(a)), it is always suppressed by the scaling of MFL.
		
	
	\begin{figure*}
		\centering
		\includegraphics[width=\linewidth]{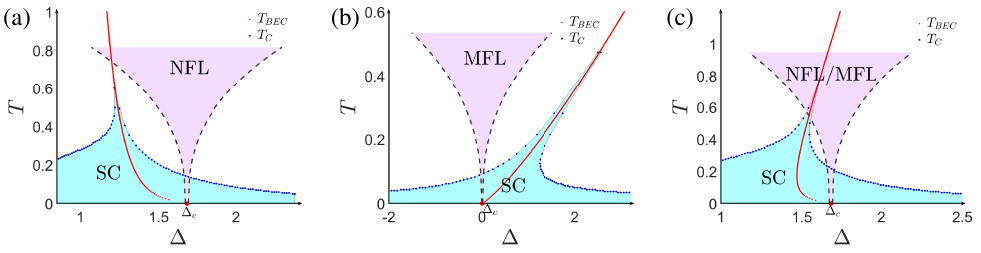}
		\caption{Finite-temperature phase diagram of model (\ref{ham}) for three different values of the SYK-liked Yukawa interactions. The phase transition temperature $T_c$ to SC is shown in blue. The transition temperature $T_{BEC}$ from the $U(1)$ phase to the BEC phase in TMI is indicated in red. Above the zero-temperature quantum critical point $\Delta_c$, a schematic depiction in purple illustrates the NFL/MFL region induced by Yukawa coupling. The parameters are: $a=-1$, $\Lambda_U=300$ and (a) $|J|/\sqrt{N}=2$, $|J'|/\sqrt{N}=0$, (b) $|J|/\sqrt{N}=0$, $|J'|/\sqrt{N}=0.6$ (c) $|J|/\sqrt{N}=2$, $|J'|/\sqrt{N}=0.2$.}
		\label{fig3}
	\end{figure*}
	
	In experiments/numerical simulations \cite{PhysRevB.95.035124, PhysRevLett.127.046601}, the fluctuations of the critical magnons couple with the Fermi surface and result in the ground state of NFL/MFL, dominating the properties of the system within a range of temperature above the critical points. On the other hand, in the low-temperature region near the critical point, a SC phase is also generated. The two effects are of comparable strength and it is not clear which dominates in previous studies. However, in our model, large-$N$ can serve as an effective controllable parameter, and vertex correlations are suppressed by $1/N$ \cite{SM}. The phase diagrams of our model (\ref{ham}) are shown in Fig. \ref{fig3}. 
	
	When only translationally invariant Yukawa interaction is present as shown in Fig. \ref{fig3}(a), the critical tuning parameter $\Delta_c(T)$ of the $U(1)$ phase and the BEC phase in TMI at finite temperature decreases as the increasing of the temperature. A SC transition is established by solving the linearized gap equation \cite{SM}. The critical temperature $T_C$ of SC increases with the increasing variance $|J|$ of the translationally invariant Yukawa coupling and satisfies $T_C\sim (|J|/\sqrt{N})^{3}$ (see the Supplemental Material \cite{SM}). The schematic NFL region is also given above the quantum critical point $\Delta_c$. When only disordered Yukawa interaction is present as shown in Fig. \ref{fig3}(b), the critical tuning parameter $\Delta_c(T)\sim T$ exhibits nearly linear growth with increasing temperature, and the mean-field exponents $\nu=1/2$ \cite{PhysRevB.14.1165}. It confirms the conclusion that the self-energy of magnons acquires thermal tuning parameter, $\Delta(T)\sim T$, when analyzing transport properties \cite{SM}. This is significantly different from considering only pervious Yukawa coupling. The critical temperature of SC, $T_C\sim \mathrm{e}^{-\sqrt{N}/|J'|}$, exponentially increases with the increasing variance $|J'|$ of the disordered SYK-liked Yukawa coupling (see the Supplemental Material \cite{SM}). When considering both interactions, as shown in Fig. \ref{fig3}(c), the critical temperature $T_{BEC}$ in the low-temperature region is determined by the interaction with spatial dependence, while in the high-temperature region, it is determined by the quenched disorder component. Therefore, the method of obtaining the temperature dependence at finite temperatures ($\gg \omega$) from the frequency $\omega$ dependence at zero temperature (i.e. $T\ll \omega$) is also applicable. In the case of $|J|=2$ and $|J'|=0.2$, the critical temperature $T_C$  of SC is dominated by the behavior of $|J|$. Fluctuations associated with  the nematic phase, particularly near a quantum critical point, can enhance the superconducting transition temperature $T_C$ \cite{doi:10.1073/pnas.1620651114, PhysRevLett.114.097001}.

	If there are multiple VHPs on the Fermi surface of materials, then the main contributions also include scattering between patches around different VHPs. As indicated by the previous scaling analysis, near the critical point, the frequency-dependent behavior of quasiparticle damping is consistent with the case when considering the scattering within the same patch. However, scattering between different patches could result in a constant contribution to the dc conductivity \cite{SM}. For real materials, symmetry is broken due to impurities or frustrations, the energy spectrum after the renormalization of higher-order diagrams always satisfies $ab^2\ne -1$ (the meaning of the parameters is explained in the Supplemental Material \cite{SM}). Therefore, the constant dc conductivity can always exist and we argue that this can contribute to the constant components in resistivity. In addition, we can also consider the superconducting pairing term $\Delta_{ij}(r)e^{i(K_1+K_2)R}\psi_{1i}^\dagger(x) \psi_{2j}^\dagger(y)$ [$R=(x+y)/2$ the position of the center of mass, $r=x-y$ the relative position, and $i$ and $j$ are flavor indices]. $\psi_1^\dagger(k)$ and $\psi_2^\dagger(k)$ are electron creation operators from two patches around different VHPs with momentum $K_1$ and $K_2$ (as shown in Fig. \ref{fig2}(c)), respectively. This PDW order with condensed momentum $K_1+K_2$ ($\ne$ the reciprocal lattice vectors $G$) will compete with the normal SC, and is believed to exist in the heavy-fermion SC \cite{radovan2003magnetic} and cuprate SC \cite{annurev-conmatphys-031119-050711}. 
	
	The scalar field in theory (\ref{ham}) is considered as magnons, which can also be viewed as order parameters or fractionalized particles in the two-dimensional materials. The Yukawa interactions (\ref{hint}) can be obtained by performing a Hubbard-Stratonovich transformation on the Hubbard term. And our method is also applicable to any type of lattice and hopping range, as long as there are VHS on the Fermi surface. The effective theory of these two-dimensional materials can also be described by Lagrangian (\ref{eff}). 
	

	We have shown that in two-dimensional materials/surface with Yukawa interactions, the VHS on the quasi-particle Fermi surface suppress regions far from the VHPs, making a significant contribution to NFL. Considering impurities can explain the $T\ln (1/T)$ specific heat and the linear temperature dependence of resistivity in strange metals. Moreover, tuning the strength of two types of interactions can lead to a crossover from NFL to MFL. In addition, we present the SC phase induced by VHS, revealing that the exponent of the SC critical temperature depends on the disorder interaction, and the power-law dependence relies on the translationally invariant one.  
	
	Our theory provides a new mechanism to explain NFL behavior and strange metal phenomena, and can be probed using real material with VHS, which are commonly present in recent popular twisted bilayer materials \cite{li2010observation, PhysRevLett.109.126801, yin2016selectively, xu2021tunable}. It is noteworthy that a recent work in Ref. \cite{PhysRevLett.130.083603} proposed a new experimental scheme using cavity quantum electrodynamics to detect NFL/MFL, and we anticipate that our theory can also be probed using similar methods. Future research will focus on determining the boundaries of the NFL/MFL region and considering the effects of fluctuations brought about beyond the Migdal-Eliashberg franework \cite{MARSIGLIO2020168102}. 
	
	\begin{acknowledgments}
		This work was supported by National Key R\&D Program of China under grants No. 2021YFA1400900, 2021YFA0718300, 2021YFA1402100, NSFC under grants Nos. 12174461, 12234012, 12334012, 52327808, Space Application System of China Manned Space Program. ECM has been partially supported by CNPq, FAPARJ and CAPES.
	\end{acknowledgments}	
		
\bibliography{paper.bib}

\begin{thebibliography}{62}%
\makeatletter
\providecommand \@ifxundefined [1]{%
 \@ifx{#1\undefined}
}%
\providecommand \@ifnum [1]{%
 \ifnum #1\expandafter \@firstoftwo
 \else \expandafter \@secondoftwo
 \fi
}%
\providecommand \@ifx [1]{%
 \ifx #1\expandafter \@firstoftwo
 \else \expandafter \@secondoftwo
 \fi
}%
\providecommand \natexlab [1]{#1}%
\providecommand \enquote  [1]{``#1''}%
\providecommand \bibnamefont  [1]{#1}%
\providecommand \bibfnamefont [1]{#1}%
\providecommand \citenamefont [1]{#1}%
\providecommand \href@noop [0]{\@secondoftwo}%
\providecommand \href [0]{\begingroup \@sanitize@url \@href}%
\providecommand \@href[1]{\@@startlink{#1}\@@href}%
\providecommand \@@href[1]{\endgroup#1\@@endlink}%
\providecommand \@sanitize@url [0]{\catcode `\\12\catcode `\$12\catcode `\&12\catcode `\#12\catcode `\^12\catcode `\_12\catcode `\%12\relax}%
\providecommand \@@startlink[1]{}%
\providecommand \@@endlink[0]{}%
\providecommand \url  [0]{\begingroup\@sanitize@url \@url }%
\providecommand \@url [1]{\endgroup\@href {#1}{\urlprefix }}%
\providecommand \urlprefix  [0]{URL }%
\providecommand \Eprint [0]{\href }%
\providecommand \doibase [0]{http://dx.doi.org/}%
\providecommand \selectlanguage [0]{\@gobble}%
\providecommand \bibinfo  [0]{\@secondoftwo}%
\providecommand \bibfield  [0]{\@secondoftwo}%
\providecommand \translation [1]{[#1]}%
\providecommand \BibitemOpen [0]{}%
\providecommand \bibitemStop [0]{}%
\providecommand \bibitemNoStop [0]{.\EOS\space}%
\providecommand \EOS [0]{\spacefactor3000\relax}%
\providecommand \BibitemShut  [1]{\csname bibitem#1\endcsname}%
\let\auto@bib@innerbib\@empty
\bibitem [{\citenamefont {Reiss}\ \emph {et~al.}(2020)\citenamefont {Reiss}, \citenamefont {Graf}, \citenamefont {Haghighirad}, \citenamefont {Knafo}, \citenamefont {Drigo}, \citenamefont {Bristow}, \citenamefont {Schofield},\ and\ \citenamefont {Coldea}}]{reiss2020quenched}%
  \BibitemOpen
  \bibfield  {author} {\bibinfo {author} {\bibfnamefont {P.}~\bibnamefont {Reiss}}, \bibinfo {author} {\bibfnamefont {D.}~\bibnamefont {Graf}}, \bibinfo {author} {\bibfnamefont {A.~A.}\ \bibnamefont {Haghighirad}}, \bibinfo {author} {\bibfnamefont {W.}~\bibnamefont {Knafo}}, \bibinfo {author} {\bibfnamefont {L.}~\bibnamefont {Drigo}}, \bibinfo {author} {\bibfnamefont {M.}~\bibnamefont {Bristow}}, \bibinfo {author} {\bibfnamefont {A.~J.}\ \bibnamefont {Schofield}}, \ and\ \bibinfo {author} {\bibfnamefont {A.~I.}\ \bibnamefont {Coldea}},\ }\href {\doibase 10.1038/s41567-019-0694-2} {\bibfield  {journal} {\bibinfo  {journal} {Nature Physics}\ }\textbf {\bibinfo {volume} {16}},\ \bibinfo {pages} {89} (\bibinfo {year} {2020})}\BibitemShut {NoStop}%
\bibitem [{\citenamefont {Lee}\ and\ \citenamefont {Phillips}(2012)}]{PhysRevB.86.245113}%
  \BibitemOpen
  \bibfield  {author} {\bibinfo {author} {\bibfnamefont {W.-C.}\ \bibnamefont {Lee}}\ and\ \bibinfo {author} {\bibfnamefont {P.~W.}\ \bibnamefont {Phillips}},\ }\href {\doibase 10.1103/PhysRevB.86.245113} {\bibfield  {journal} {\bibinfo  {journal} {Phys. Rev. B}\ }\textbf {\bibinfo {volume} {86}},\ \bibinfo {pages} {245113} (\bibinfo {year} {2012})}\BibitemShut {NoStop}%
\bibitem [{\citenamefont {Chen}\ \emph {et~al.}(2019{\natexlab{a}})\citenamefont {Chen}, \citenamefont {Wang}, \citenamefont {Li}, \citenamefont {Feng}, \citenamefont {Dai}, \citenamefont {Xu},\ and\ \citenamefont {Si}}]{chen2019heavy}%
  \BibitemOpen
  \bibfield  {author} {\bibinfo {author} {\bibfnamefont {J.}~\bibnamefont {Chen}}, \bibinfo {author} {\bibfnamefont {Z.}~\bibnamefont {Wang}}, \bibinfo {author} {\bibfnamefont {Y.}~\bibnamefont {Li}}, \bibinfo {author} {\bibfnamefont {C.}~\bibnamefont {Feng}}, \bibinfo {author} {\bibfnamefont {J.}~\bibnamefont {Dai}}, \bibinfo {author} {\bibfnamefont {Z.}~\bibnamefont {Xu}}, \ and\ \bibinfo {author} {\bibfnamefont {Q.}~\bibnamefont {Si}},\ }\href {\doibase 10.1038/s41598-019-48662-8} {\bibfield  {journal} {\bibinfo  {journal} {Scientific Reports}\ }\textbf {\bibinfo {volume} {9}},\ \bibinfo {pages} {12307} (\bibinfo {year} {2019}{\natexlab{a}})}\BibitemShut {NoStop}%
\bibitem [{\citenamefont {Seiro}\ \emph {et~al.}(2018)\citenamefont {Seiro}, \citenamefont {Jiao}, \citenamefont {Kirchner}, \citenamefont {Hartmann}, \citenamefont {Friedemann}, \citenamefont {Krellner}, \citenamefont {Geibel}, \citenamefont {Si}, \citenamefont {Steglich},\ and\ \citenamefont {Wirth}}]{seiro2018evolution}%
  \BibitemOpen
  \bibfield  {author} {\bibinfo {author} {\bibfnamefont {S.}~\bibnamefont {Seiro}}, \bibinfo {author} {\bibfnamefont {L.}~\bibnamefont {Jiao}}, \bibinfo {author} {\bibfnamefont {S.}~\bibnamefont {Kirchner}}, \bibinfo {author} {\bibfnamefont {S.}~\bibnamefont {Hartmann}}, \bibinfo {author} {\bibfnamefont {S.}~\bibnamefont {Friedemann}}, \bibinfo {author} {\bibfnamefont {C.}~\bibnamefont {Krellner}}, \bibinfo {author} {\bibfnamefont {C.}~\bibnamefont {Geibel}}, \bibinfo {author} {\bibfnamefont {Q.}~\bibnamefont {Si}}, \bibinfo {author} {\bibfnamefont {F.}~\bibnamefont {Steglich}}, \ and\ \bibinfo {author} {\bibfnamefont {S.}~\bibnamefont {Wirth}},\ }\href {\doibase 10.1038/s41467-018-05801-5} {\bibfield  {journal} {\bibinfo  {journal} {Nature Communications}\ }\textbf {\bibinfo {volume} {9}},\ \bibinfo {pages} {3324} (\bibinfo {year} {2018})}\BibitemShut {NoStop}%
\bibitem [{\citenamefont {Xu}\ \emph {et~al.}(2020)\citenamefont {Xu}, \citenamefont {Wu}, \citenamefont {Jian},\ and\ \citenamefont {Xu}}]{PhysRevB.101.205426}%
  \BibitemOpen
  \bibfield  {author} {\bibinfo {author} {\bibfnamefont {Y.}~\bibnamefont {Xu}}, \bibinfo {author} {\bibfnamefont {X.-C.}\ \bibnamefont {Wu}}, \bibinfo {author} {\bibfnamefont {C.-M.}\ \bibnamefont {Jian}}, \ and\ \bibinfo {author} {\bibfnamefont {C.}~\bibnamefont {Xu}},\ }\href {\doibase 10.1103/PhysRevB.101.205426} {\bibfield  {journal} {\bibinfo  {journal} {Phys. Rev. B}\ }\textbf {\bibinfo {volume} {101}},\ \bibinfo {pages} {205426} (\bibinfo {year} {2020})}\BibitemShut {NoStop}%
\bibitem [{\citenamefont {Wang}\ \emph {et~al.}(2022)\citenamefont {Wang}, \citenamefont {Yu}, \citenamefont {Kwan}, \citenamefont {Jia}, \citenamefont {Lei}, \citenamefont {Klemenz}, \citenamefont {Cevallos}, \citenamefont {Singha}, \citenamefont {Devakul}, \citenamefont {Watanabe} \emph {et~al.}}]{wang2022one}%
  \BibitemOpen
  \bibfield  {author} {\bibinfo {author} {\bibfnamefont {P.}~\bibnamefont {Wang}}, \bibinfo {author} {\bibfnamefont {G.}~\bibnamefont {Yu}}, \bibinfo {author} {\bibfnamefont {Y.~H.}\ \bibnamefont {Kwan}}, \bibinfo {author} {\bibfnamefont {Y.}~\bibnamefont {Jia}}, \bibinfo {author} {\bibfnamefont {S.}~\bibnamefont {Lei}}, \bibinfo {author} {\bibfnamefont {S.}~\bibnamefont {Klemenz}}, \bibinfo {author} {\bibfnamefont {F.~A.}\ \bibnamefont {Cevallos}}, \bibinfo {author} {\bibfnamefont {R.}~\bibnamefont {Singha}}, \bibinfo {author} {\bibfnamefont {T.}~\bibnamefont {Devakul}}, \bibinfo {author} {\bibfnamefont {K.}~\bibnamefont {Watanabe}},  \emph {et~al.},\ }\href {\doibase 10.1038/s41586-022-04514-6} {\bibfield  {journal} {\bibinfo  {journal} {Nature}\ }\textbf {\bibinfo {volume} {605}},\ \bibinfo {pages} {57} (\bibinfo {year} {2022})}\BibitemShut {NoStop}%
\bibitem [{\citenamefont {Cao}\ \emph {et~al.}(2020)\citenamefont {Cao}, \citenamefont {Chowdhury}, \citenamefont {Rodan-Legrain}, \citenamefont {Rubies-Bigorda}, \citenamefont {Watanabe}, \citenamefont {Taniguchi}, \citenamefont {Senthil},\ and\ \citenamefont {Jarillo-Herrero}}]{PhysRevLett.124.076801}%
  \BibitemOpen
  \bibfield  {author} {\bibinfo {author} {\bibfnamefont {Y.}~\bibnamefont {Cao}}, \bibinfo {author} {\bibfnamefont {D.}~\bibnamefont {Chowdhury}}, \bibinfo {author} {\bibfnamefont {D.}~\bibnamefont {Rodan-Legrain}}, \bibinfo {author} {\bibfnamefont {O.}~\bibnamefont {Rubies-Bigorda}}, \bibinfo {author} {\bibfnamefont {K.}~\bibnamefont {Watanabe}}, \bibinfo {author} {\bibfnamefont {T.}~\bibnamefont {Taniguchi}}, \bibinfo {author} {\bibfnamefont {T.}~\bibnamefont {Senthil}}, \ and\ \bibinfo {author} {\bibfnamefont {P.}~\bibnamefont {Jarillo-Herrero}},\ }\href {\doibase 10.1103/PhysRevLett.124.076801} {\bibfield  {journal} {\bibinfo  {journal} {Phys. Rev. Lett.}\ }\textbf {\bibinfo {volume} {124}},\ \bibinfo {pages} {076801} (\bibinfo {year} {2020})}\BibitemShut {NoStop}%
\bibitem [{\citenamefont {Varma}\ \emph {et~al.}(1989)\citenamefont {Varma}, \citenamefont {Littlewood}, \citenamefont {Schmitt-Rink}, \citenamefont {Abrahams},\ and\ \citenamefont {Ruckenstein}}]{PhysRevLett.63.1996}%
  \BibitemOpen
  \bibfield  {author} {\bibinfo {author} {\bibfnamefont {C.~M.}\ \bibnamefont {Varma}}, \bibinfo {author} {\bibfnamefont {P.~B.}\ \bibnamefont {Littlewood}}, \bibinfo {author} {\bibfnamefont {S.}~\bibnamefont {Schmitt-Rink}}, \bibinfo {author} {\bibfnamefont {E.}~\bibnamefont {Abrahams}}, \ and\ \bibinfo {author} {\bibfnamefont {A.~E.}\ \bibnamefont {Ruckenstein}},\ }\href {\doibase 10.1103/PhysRevLett.63.1996} {\bibfield  {journal} {\bibinfo  {journal} {Phys. Rev. Lett.}\ }\textbf {\bibinfo {volume} {63}},\ \bibinfo {pages} {1996} (\bibinfo {year} {1989})}\BibitemShut {NoStop}%
\bibitem [{\citenamefont {Kakehashi}\ and\ \citenamefont {Fulde}(2005)}]{PhysRevLett.94.156401}%
  \BibitemOpen
  \bibfield  {author} {\bibinfo {author} {\bibfnamefont {Y.}~\bibnamefont {Kakehashi}}\ and\ \bibinfo {author} {\bibfnamefont {P.}~\bibnamefont {Fulde}},\ }\href {\doibase 10.1103/PhysRevLett.94.156401} {\bibfield  {journal} {\bibinfo  {journal} {Phys. Rev. Lett.}\ }\textbf {\bibinfo {volume} {94}},\ \bibinfo {pages} {156401} (\bibinfo {year} {2005})}\BibitemShut {NoStop}%
\bibitem [{CFM()}]{CFM}%
  \BibitemOpen
  \href@noop {} {}\bibinfo {note} {At the quantum critical point, electronic excitations exhibit a well-defined Fermi surface but no sharp Landau quasiparticles.}\BibitemShut {Stop}%
\bibitem [{\citenamefont {Patel}\ \emph {et~al.}(2023)\citenamefont {Patel}, \citenamefont {Guo}, \citenamefont {Esterlis},\ and\ \citenamefont {Sachdev}}]{science.abq6011}%
  \BibitemOpen
  \bibfield  {author} {\bibinfo {author} {\bibfnamefont {A.~A.}\ \bibnamefont {Patel}}, \bibinfo {author} {\bibfnamefont {H.}~\bibnamefont {Guo}}, \bibinfo {author} {\bibfnamefont {I.}~\bibnamefont {Esterlis}}, \ and\ \bibinfo {author} {\bibfnamefont {S.}~\bibnamefont {Sachdev}},\ }\href {\doibase 10.1126/science.abq6011} {\bibfield  {journal} {\bibinfo  {journal} {Science}\ }\textbf {\bibinfo {volume} {381}},\ \bibinfo {pages} {790} (\bibinfo {year} {2023})}\BibitemShut {NoStop}%
\bibitem [{\citenamefont {Esterlis}\ \emph {et~al.}(2021)\citenamefont {Esterlis}, \citenamefont {Guo}, \citenamefont {Patel},\ and\ \citenamefont {Sachdev}}]{PhysRevB.103.235129}%
  \BibitemOpen
  \bibfield  {author} {\bibinfo {author} {\bibfnamefont {I.}~\bibnamefont {Esterlis}}, \bibinfo {author} {\bibfnamefont {H.}~\bibnamefont {Guo}}, \bibinfo {author} {\bibfnamefont {A.~A.}\ \bibnamefont {Patel}}, \ and\ \bibinfo {author} {\bibfnamefont {S.}~\bibnamefont {Sachdev}},\ }\href {\doibase 10.1103/PhysRevB.103.235129} {\bibfield  {journal} {\bibinfo  {journal} {Phys. Rev. B}\ }\textbf {\bibinfo {volume} {103}},\ \bibinfo {pages} {235129} (\bibinfo {year} {2021})}\BibitemShut {NoStop}%
\bibitem [{\citenamefont {Guo}\ \emph {et~al.}(2022)\citenamefont {Guo}, \citenamefont {Patel}, \citenamefont {Esterlis},\ and\ \citenamefont {Sachdev}}]{PhysRevB.106.115151}%
  \BibitemOpen
  \bibfield  {author} {\bibinfo {author} {\bibfnamefont {H.}~\bibnamefont {Guo}}, \bibinfo {author} {\bibfnamefont {A.~A.}\ \bibnamefont {Patel}}, \bibinfo {author} {\bibfnamefont {I.}~\bibnamefont {Esterlis}}, \ and\ \bibinfo {author} {\bibfnamefont {S.}~\bibnamefont {Sachdev}},\ }\href {\doibase 10.1103/PhysRevB.106.115151} {\bibfield  {journal} {\bibinfo  {journal} {Phys. Rev. B}\ }\textbf {\bibinfo {volume} {106}},\ \bibinfo {pages} {115151} (\bibinfo {year} {2022})}\BibitemShut {NoStop}%
\bibitem [{\citenamefont {Lee}(2009)}]{physRevB.80.165102}%
  \BibitemOpen
  \bibfield  {author} {\bibinfo {author} {\bibfnamefont {S.-S.}\ \bibnamefont {Lee}},\ }\href {\doibase 10.1103/PhysRevB.80.165102} {\bibfield  {journal} {\bibinfo  {journal} {Phys. Rev. B}\ }\textbf {\bibinfo {volume} {80}},\ \bibinfo {pages} {165102} (\bibinfo {year} {2009})}\BibitemShut {NoStop}%
\bibitem [{\citenamefont {Holder}\ and\ \citenamefont {Metzner}(2014)}]{PhysRevB.90.161106}%
  \BibitemOpen
  \bibfield  {author} {\bibinfo {author} {\bibfnamefont {T.}~\bibnamefont {Holder}}\ and\ \bibinfo {author} {\bibfnamefont {W.}~\bibnamefont {Metzner}},\ }\href {\doibase 10.1103/PhysRevB.90.161106} {\bibfield  {journal} {\bibinfo  {journal} {Phys. Rev. B}\ }\textbf {\bibinfo {volume} {90}},\ \bibinfo {pages} {161106} (\bibinfo {year} {2014})}\BibitemShut {NoStop}%
\bibitem [{\citenamefont {Rao}\ and\ \citenamefont {Piazza}(2023)}]{PhysRevLett.130.083603}%
  \BibitemOpen
  \bibfield  {author} {\bibinfo {author} {\bibfnamefont {P.}~\bibnamefont {Rao}}\ and\ \bibinfo {author} {\bibfnamefont {F.}~\bibnamefont {Piazza}},\ }\href {\doibase 10.1103/PhysRevLett.130.083603} {\bibfield  {journal} {\bibinfo  {journal} {Phys. Rev. Lett.}\ }\textbf {\bibinfo {volume} {130}},\ \bibinfo {pages} {083603} (\bibinfo {year} {2023})}\BibitemShut {NoStop}%
\bibitem [{\citenamefont {Polshyn}\ \emph {et~al.}(2019)\citenamefont {Polshyn}, \citenamefont {Yankowitz}, \citenamefont {Chen}, \citenamefont {Zhang}, \citenamefont {Watanabe}, \citenamefont {Taniguchi}, \citenamefont {Dean},\ and\ \citenamefont {Young}}]{polshyn2019large}%
  \BibitemOpen
  \bibfield  {author} {\bibinfo {author} {\bibfnamefont {H.}~\bibnamefont {Polshyn}}, \bibinfo {author} {\bibfnamefont {M.}~\bibnamefont {Yankowitz}}, \bibinfo {author} {\bibfnamefont {S.}~\bibnamefont {Chen}}, \bibinfo {author} {\bibfnamefont {Y.}~\bibnamefont {Zhang}}, \bibinfo {author} {\bibfnamefont {K.}~\bibnamefont {Watanabe}}, \bibinfo {author} {\bibfnamefont {T.}~\bibnamefont {Taniguchi}}, \bibinfo {author} {\bibfnamefont {C.~R.}\ \bibnamefont {Dean}}, \ and\ \bibinfo {author} {\bibfnamefont {A.~F.}\ \bibnamefont {Young}},\ }\href {\doibase 10.1038/s41567-019-0596-3} {\bibfield  {journal} {\bibinfo  {journal} {Nature Physics}\ }\textbf {\bibinfo {volume} {15}},\ \bibinfo {pages} {1011} (\bibinfo {year} {2019})}\BibitemShut {NoStop}%
\bibitem [{\citenamefont {Mousatov}\ \emph {et~al.}(2020)\citenamefont {Mousatov}, \citenamefont {Berg},\ and\ \citenamefont {Hartnoll}}]{mousatov2020theory}%
  \BibitemOpen
  \bibfield  {author} {\bibinfo {author} {\bibfnamefont {C.~H.}\ \bibnamefont {Mousatov}}, \bibinfo {author} {\bibfnamefont {E.}~\bibnamefont {Berg}}, \ and\ \bibinfo {author} {\bibfnamefont {S.~A.}\ \bibnamefont {Hartnoll}},\ }\href {\doibase 10.1073/pnas.1915224117} {\bibfield  {journal} {\bibinfo  {journal} {Proceedings of the National Academy of Sciences}\ }\textbf {\bibinfo {volume} {117}},\ \bibinfo {pages} {2852} (\bibinfo {year} {2020})}\BibitemShut {NoStop}%
\bibitem [{\citenamefont {Xu}\ \emph {et~al.}(2021{\natexlab{a}})\citenamefont {Xu}, \citenamefont {Herman}, \citenamefont {Granata}, \citenamefont {Destraz}, \citenamefont {Das}, \citenamefont {Vonka}, \citenamefont {Gerber}, \citenamefont {Spring}, \citenamefont {Gibert}, \citenamefont {Schilling} \emph {et~al.}}]{xu2021magnetotransport}%
  \BibitemOpen
  \bibfield  {author} {\bibinfo {author} {\bibfnamefont {Y.}~\bibnamefont {Xu}}, \bibinfo {author} {\bibfnamefont {F.}~\bibnamefont {Herman}}, \bibinfo {author} {\bibfnamefont {V.}~\bibnamefont {Granata}}, \bibinfo {author} {\bibfnamefont {D.}~\bibnamefont {Destraz}}, \bibinfo {author} {\bibfnamefont {L.}~\bibnamefont {Das}}, \bibinfo {author} {\bibfnamefont {J.}~\bibnamefont {Vonka}}, \bibinfo {author} {\bibfnamefont {S.}~\bibnamefont {Gerber}}, \bibinfo {author} {\bibfnamefont {J.}~\bibnamefont {Spring}}, \bibinfo {author} {\bibfnamefont {M.}~\bibnamefont {Gibert}}, \bibinfo {author} {\bibfnamefont {A.}~\bibnamefont {Schilling}},  \emph {et~al.},\ }\href {\doibase 10.1038/s42005-020-00504-0} {\bibfield  {journal} {\bibinfo  {journal} {Communications Physics}\ }\textbf {\bibinfo {volume} {4}},\ \bibinfo {pages} {1} (\bibinfo {year} {2021}{\natexlab{a}})}\BibitemShut {NoStop}%
\bibitem [{\citenamefont {Wang}\ \emph {et~al.}(2020)\citenamefont {Wang}, \citenamefont {Shih}, \citenamefont {Ghiotto}, \citenamefont {Xian}, \citenamefont {Rhodes}, \citenamefont {Tan}, \citenamefont {Claassen}, \citenamefont {Kennes}, \citenamefont {Bai}, \citenamefont {Kim} \emph {et~al.}}]{wang2020correlated}%
  \BibitemOpen
  \bibfield  {author} {\bibinfo {author} {\bibfnamefont {L.}~\bibnamefont {Wang}}, \bibinfo {author} {\bibfnamefont {E.-M.}\ \bibnamefont {Shih}}, \bibinfo {author} {\bibfnamefont {A.}~\bibnamefont {Ghiotto}}, \bibinfo {author} {\bibfnamefont {L.}~\bibnamefont {Xian}}, \bibinfo {author} {\bibfnamefont {D.~A.}\ \bibnamefont {Rhodes}}, \bibinfo {author} {\bibfnamefont {C.}~\bibnamefont {Tan}}, \bibinfo {author} {\bibfnamefont {M.}~\bibnamefont {Claassen}}, \bibinfo {author} {\bibfnamefont {D.~M.}\ \bibnamefont {Kennes}}, \bibinfo {author} {\bibfnamefont {Y.}~\bibnamefont {Bai}}, \bibinfo {author} {\bibfnamefont {B.}~\bibnamefont {Kim}},  \emph {et~al.},\ }\href {\doibase 10.1038/s41563-020-0708-6} {\bibfield  {journal} {\bibinfo  {journal} {Nature materials}\ }\textbf {\bibinfo {volume} {19}},\ \bibinfo {pages} {861} (\bibinfo {year} {2020})}\BibitemShut {NoStop}%
\bibitem [{\citenamefont {Yudin}\ \emph {et~al.}(2014)\citenamefont {Yudin}, \citenamefont {Hirschmeier}, \citenamefont {Hafermann}, \citenamefont {Eriksson}, \citenamefont {Lichtenstein},\ and\ \citenamefont {Katsnelson}}]{PhysRevLett.112.070403}%
  \BibitemOpen
  \bibfield  {author} {\bibinfo {author} {\bibfnamefont {D.}~\bibnamefont {Yudin}}, \bibinfo {author} {\bibfnamefont {D.}~\bibnamefont {Hirschmeier}}, \bibinfo {author} {\bibfnamefont {H.}~\bibnamefont {Hafermann}}, \bibinfo {author} {\bibfnamefont {O.}~\bibnamefont {Eriksson}}, \bibinfo {author} {\bibfnamefont {A.~I.}\ \bibnamefont {Lichtenstein}}, \ and\ \bibinfo {author} {\bibfnamefont {M.~I.}\ \bibnamefont {Katsnelson}},\ }\href {\doibase 10.1103/PhysRevLett.112.070403} {\bibfield  {journal} {\bibinfo  {journal} {Phys. Rev. Lett.}\ }\textbf {\bibinfo {volume} {112}},\ \bibinfo {pages} {070403} (\bibinfo {year} {2014})}\BibitemShut {NoStop}%
\bibitem [{\citenamefont {Karp}\ \emph {et~al.}(2020)\citenamefont {Karp}, \citenamefont {Bramberger}, \citenamefont {Grundner}, \citenamefont {Schollw\"ock}, \citenamefont {Millis},\ and\ \citenamefont {Zingl}}]{PhysRevLett.125.166401}%
  \BibitemOpen
  \bibfield  {author} {\bibinfo {author} {\bibfnamefont {J.}~\bibnamefont {Karp}}, \bibinfo {author} {\bibfnamefont {M.}~\bibnamefont {Bramberger}}, \bibinfo {author} {\bibfnamefont {M.}~\bibnamefont {Grundner}}, \bibinfo {author} {\bibfnamefont {U.}~\bibnamefont {Schollw\"ock}}, \bibinfo {author} {\bibfnamefont {A.~J.}\ \bibnamefont {Millis}}, \ and\ \bibinfo {author} {\bibfnamefont {M.}~\bibnamefont {Zingl}},\ }\href {\doibase 10.1103/PhysRevLett.125.166401} {\bibfield  {journal} {\bibinfo  {journal} {Phys. Rev. Lett.}\ }\textbf {\bibinfo {volume} {125}},\ \bibinfo {pages} {166401} (\bibinfo {year} {2020})}\BibitemShut {NoStop}%
\bibitem [{\citenamefont {Kang}\ \emph {et~al.}(2022)\citenamefont {Kang}, \citenamefont {Fang}, \citenamefont {Kim}, \citenamefont {Ortiz}, \citenamefont {Ryu}, \citenamefont {Kim}, \citenamefont {Yoo}, \citenamefont {Sangiovanni}, \citenamefont {Di~Sante}, \citenamefont {Park} \emph {et~al.}}]{kang2022twofold}%
  \BibitemOpen
  \bibfield  {author} {\bibinfo {author} {\bibfnamefont {M.}~\bibnamefont {Kang}}, \bibinfo {author} {\bibfnamefont {S.}~\bibnamefont {Fang}}, \bibinfo {author} {\bibfnamefont {J.-K.}\ \bibnamefont {Kim}}, \bibinfo {author} {\bibfnamefont {B.~R.}\ \bibnamefont {Ortiz}}, \bibinfo {author} {\bibfnamefont {S.~H.}\ \bibnamefont {Ryu}}, \bibinfo {author} {\bibfnamefont {J.}~\bibnamefont {Kim}}, \bibinfo {author} {\bibfnamefont {J.}~\bibnamefont {Yoo}}, \bibinfo {author} {\bibfnamefont {G.}~\bibnamefont {Sangiovanni}}, \bibinfo {author} {\bibfnamefont {D.}~\bibnamefont {Di~Sante}}, \bibinfo {author} {\bibfnamefont {B.-G.}\ \bibnamefont {Park}},  \emph {et~al.},\ }\href {\doibase 10.1038/s41567-021-01451-5} {\bibfield  {journal} {\bibinfo  {journal} {Nature Physics}\ }\textbf {\bibinfo {volume} {18}},\ \bibinfo {pages} {301} (\bibinfo {year} {2022})}\BibitemShut {NoStop}%
\bibitem [{\citenamefont {Wu}\ \emph {et~al.}(2021)\citenamefont {Wu}, \citenamefont {Zhang}, \citenamefont {Watanabe}, \citenamefont {Taniguchi},\ and\ \citenamefont {Andrei}}]{wu2021chern}%
  \BibitemOpen
  \bibfield  {author} {\bibinfo {author} {\bibfnamefont {S.}~\bibnamefont {Wu}}, \bibinfo {author} {\bibfnamefont {Z.}~\bibnamefont {Zhang}}, \bibinfo {author} {\bibfnamefont {K.}~\bibnamefont {Watanabe}}, \bibinfo {author} {\bibfnamefont {T.}~\bibnamefont {Taniguchi}}, \ and\ \bibinfo {author} {\bibfnamefont {E.~Y.}\ \bibnamefont {Andrei}},\ }\href {\doibase 10.1038/s41563-020-00911-2} {\bibfield  {journal} {\bibinfo  {journal} {Nature materials}\ }\textbf {\bibinfo {volume} {20}},\ \bibinfo {pages} {488} (\bibinfo {year} {2021})}\BibitemShut {NoStop}%
\bibitem [{\citenamefont {Luo}\ \emph {et~al.}(2023)\citenamefont {Luo}, \citenamefont {Han}, \citenamefont {Liu}, \citenamefont {Chen}, \citenamefont {Huang}, \citenamefont {Huai}, \citenamefont {Li}, \citenamefont {Wang}, \citenamefont {Shen}, \citenamefont {Ding} \emph {et~al.}}]{luo2023unique}%
  \BibitemOpen
  \bibfield  {author} {\bibinfo {author} {\bibfnamefont {Y.}~\bibnamefont {Luo}}, \bibinfo {author} {\bibfnamefont {Y.}~\bibnamefont {Han}}, \bibinfo {author} {\bibfnamefont {J.}~\bibnamefont {Liu}}, \bibinfo {author} {\bibfnamefont {H.}~\bibnamefont {Chen}}, \bibinfo {author} {\bibfnamefont {Z.}~\bibnamefont {Huang}}, \bibinfo {author} {\bibfnamefont {L.}~\bibnamefont {Huai}}, \bibinfo {author} {\bibfnamefont {H.}~\bibnamefont {Li}}, \bibinfo {author} {\bibfnamefont {B.}~\bibnamefont {Wang}}, \bibinfo {author} {\bibfnamefont {J.}~\bibnamefont {Shen}}, \bibinfo {author} {\bibfnamefont {S.}~\bibnamefont {Ding}},  \emph {et~al.},\ }\href {\doibase 10.1038/s41467-023-39500-7} {\bibfield  {journal} {\bibinfo  {journal} {Nature Communications}\ }\textbf {\bibinfo {volume} {14}},\ \bibinfo {pages} {3819} (\bibinfo {year} {2023})}\BibitemShut {NoStop}%
\bibitem [{\citenamefont {Sano}\ \emph {et~al.}(2016)\citenamefont {Sano}, \citenamefont {Koretsune}, \citenamefont {Tadano}, \citenamefont {Akashi},\ and\ \citenamefont {Arita}}]{PhysRevB.93.094525}%
  \BibitemOpen
  \bibfield  {author} {\bibinfo {author} {\bibfnamefont {W.}~\bibnamefont {Sano}}, \bibinfo {author} {\bibfnamefont {T.}~\bibnamefont {Koretsune}}, \bibinfo {author} {\bibfnamefont {T.}~\bibnamefont {Tadano}}, \bibinfo {author} {\bibfnamefont {R.}~\bibnamefont {Akashi}}, \ and\ \bibinfo {author} {\bibfnamefont {R.}~\bibnamefont {Arita}},\ }\href {\doibase 10.1103/PhysRevB.93.094525} {\bibfield  {journal} {\bibinfo  {journal} {Phys. Rev. B}\ }\textbf {\bibinfo {volume} {93}},\ \bibinfo {pages} {094525} (\bibinfo {year} {2016})}\BibitemShut {NoStop}%
\bibitem [{\citenamefont {Irkhin}\ \emph {et~al.}(2002)\citenamefont {Irkhin}, \citenamefont {Katanin},\ and\ \citenamefont {Katsnelson}}]{PhysRevLett.89.076401}%
  \BibitemOpen
  \bibfield  {author} {\bibinfo {author} {\bibfnamefont {V.~Y.}\ \bibnamefont {Irkhin}}, \bibinfo {author} {\bibfnamefont {A.~A.}\ \bibnamefont {Katanin}}, \ and\ \bibinfo {author} {\bibfnamefont {M.~I.}\ \bibnamefont {Katsnelson}},\ }\href {\doibase 10.1103/PhysRevLett.89.076401} {\bibfield  {journal} {\bibinfo  {journal} {Phys. Rev. Lett.}\ }\textbf {\bibinfo {volume} {89}},\ \bibinfo {pages} {076401} (\bibinfo {year} {2002})}\BibitemShut {NoStop}%
\bibitem [{\citenamefont {Wu}\ \emph {et~al.}(2023{\natexlab{a}})\citenamefont {Wu}, \citenamefont {Wu},\ and\ \citenamefont {Yao}}]{PhysRevLett.130.126001}%
  \BibitemOpen
  \bibfield  {author} {\bibinfo {author} {\bibfnamefont {Y.-M.}\ \bibnamefont {Wu}}, \bibinfo {author} {\bibfnamefont {Z.}~\bibnamefont {Wu}}, \ and\ \bibinfo {author} {\bibfnamefont {H.}~\bibnamefont {Yao}},\ }\href {\doibase 10.1103/PhysRevLett.130.126001} {\bibfield  {journal} {\bibinfo  {journal} {Phys. Rev. Lett.}\ }\textbf {\bibinfo {volume} {130}},\ \bibinfo {pages} {126001} (\bibinfo {year} {2023}{\natexlab{a}})}\BibitemShut {NoStop}%
\bibitem [{\citenamefont {Castro}\ \emph {et~al.}(2023)\citenamefont {Castro}, \citenamefont {Shaffer}, \citenamefont {Wu},\ and\ \citenamefont {Santos}}]{PhysRevLett.131.026601}%
  \BibitemOpen
  \bibfield  {author} {\bibinfo {author} {\bibfnamefont {P.}~\bibnamefont {Castro}}, \bibinfo {author} {\bibfnamefont {D.}~\bibnamefont {Shaffer}}, \bibinfo {author} {\bibfnamefont {Y.-M.}\ \bibnamefont {Wu}}, \ and\ \bibinfo {author} {\bibfnamefont {L.~H.}\ \bibnamefont {Santos}},\ }\href {\doibase 10.1103/PhysRevLett.131.026601} {\bibfield  {journal} {\bibinfo  {journal} {Phys. Rev. Lett.}\ }\textbf {\bibinfo {volume} {131}},\ \bibinfo {pages} {026601} (\bibinfo {year} {2023})}\BibitemShut {NoStop}%
\bibitem [{\citenamefont {Sachdev}\ and\ \citenamefont {Ye}(1993)}]{PhysRevLett.70.3339}%
  \BibitemOpen
  \bibfield  {author} {\bibinfo {author} {\bibfnamefont {S.}~\bibnamefont {Sachdev}}\ and\ \bibinfo {author} {\bibfnamefont {J.}~\bibnamefont {Ye}},\ }\href {\doibase 10.1103/PhysRevLett.70.3339} {\bibfield  {journal} {\bibinfo  {journal} {Phys. Rev. Lett.}\ }\textbf {\bibinfo {volume} {70}},\ \bibinfo {pages} {3339} (\bibinfo {year} {1993})}\BibitemShut {NoStop}%
\bibitem [{\citenamefont {Wang}(2020)}]{PhysRevLett.124.017002}%
  \BibitemOpen
  \bibfield  {author} {\bibinfo {author} {\bibfnamefont {Y.}~\bibnamefont {Wang}},\ }\href {\doibase 10.1103/PhysRevLett.124.017002} {\bibfield  {journal} {\bibinfo  {journal} {Phys. Rev. Lett.}\ }\textbf {\bibinfo {volume} {124}},\ \bibinfo {pages} {017002} (\bibinfo {year} {2020})}\BibitemShut {NoStop}%
\bibitem [{\citenamefont {Chowdhury}\ \emph {et~al.}(2018)\citenamefont {Chowdhury}, \citenamefont {Werman}, \citenamefont {Berg},\ and\ \citenamefont {Senthil}}]{PhysRevX.8.031024}%
  \BibitemOpen
  \bibfield  {author} {\bibinfo {author} {\bibfnamefont {D.}~\bibnamefont {Chowdhury}}, \bibinfo {author} {\bibfnamefont {Y.}~\bibnamefont {Werman}}, \bibinfo {author} {\bibfnamefont {E.}~\bibnamefont {Berg}}, \ and\ \bibinfo {author} {\bibfnamefont {T.}~\bibnamefont {Senthil}},\ }\href {\doibase 10.1103/PhysRevX.8.031024} {\bibfield  {journal} {\bibinfo  {journal} {Phys. Rev. X}\ }\textbf {\bibinfo {volume} {8}},\ \bibinfo {pages} {031024} (\bibinfo {year} {2018})}\BibitemShut {NoStop}%
\bibitem [{\citenamefont {Wu}\ \emph {et~al.}(2023{\natexlab{b}})\citenamefont {Wu}, \citenamefont {Nosov}, \citenamefont {Patel},\ and\ \citenamefont {Raghu}}]{PhysRevLett.130.026001}%
  \BibitemOpen
  \bibfield  {author} {\bibinfo {author} {\bibfnamefont {Y.-M.}\ \bibnamefont {Wu}}, \bibinfo {author} {\bibfnamefont {P.~A.}\ \bibnamefont {Nosov}}, \bibinfo {author} {\bibfnamefont {A.~A.}\ \bibnamefont {Patel}}, \ and\ \bibinfo {author} {\bibfnamefont {S.}~\bibnamefont {Raghu}},\ }\href {\doibase 10.1103/PhysRevLett.130.026001} {\bibfield  {journal} {\bibinfo  {journal} {Phys. Rev. Lett.}\ }\textbf {\bibinfo {volume} {130}},\ \bibinfo {pages} {026001} (\bibinfo {year} {2023}{\natexlab{b}})}\BibitemShut {NoStop}%
\bibitem [{\citenamefont {Sachdev}(2015)}]{PhysRevX.5.041025}%
  \BibitemOpen
  \bibfield  {author} {\bibinfo {author} {\bibfnamefont {S.}~\bibnamefont {Sachdev}},\ }\href {\doibase 10.1103/PhysRevX.5.041025} {\bibfield  {journal} {\bibinfo  {journal} {Phys. Rev. X}\ }\textbf {\bibinfo {volume} {5}},\ \bibinfo {pages} {041025} (\bibinfo {year} {2015})}\BibitemShut {NoStop}%
\bibitem [{\citenamefont {Chen}\ \emph {et~al.}(2019{\natexlab{b}})\citenamefont {Chen}, \citenamefont {Hashimoto}, \citenamefont {He}, \citenamefont {Song}, \citenamefont {Xu}, \citenamefont {He}, \citenamefont {Devereaux}, \citenamefont {Eisaki}, \citenamefont {Lu}, \citenamefont {Zaanen} \emph {et~al.}}]{chen2019incoherent}%
  \BibitemOpen
  \bibfield  {author} {\bibinfo {author} {\bibfnamefont {S.-D.}\ \bibnamefont {Chen}}, \bibinfo {author} {\bibfnamefont {M.}~\bibnamefont {Hashimoto}}, \bibinfo {author} {\bibfnamefont {Y.}~\bibnamefont {He}}, \bibinfo {author} {\bibfnamefont {D.}~\bibnamefont {Song}}, \bibinfo {author} {\bibfnamefont {K.-J.}\ \bibnamefont {Xu}}, \bibinfo {author} {\bibfnamefont {J.-F.}\ \bibnamefont {He}}, \bibinfo {author} {\bibfnamefont {T.~P.}\ \bibnamefont {Devereaux}}, \bibinfo {author} {\bibfnamefont {H.}~\bibnamefont {Eisaki}}, \bibinfo {author} {\bibfnamefont {D.-H.}\ \bibnamefont {Lu}}, \bibinfo {author} {\bibfnamefont {J.}~\bibnamefont {Zaanen}},  \emph {et~al.},\ }\href {\doibase 10.1126/science.aaw8850} {\bibfield  {journal} {\bibinfo  {journal} {Science}\ }\textbf {\bibinfo {volume} {366}},\ \bibinfo {pages} {1099} (\bibinfo {year} {2019}{\natexlab{b}})}\BibitemShut {NoStop}%
\bibitem [{\citenamefont {Chowdhury}\ \emph {et~al.}(2022)\citenamefont {Chowdhury}, \citenamefont {Georges}, \citenamefont {Parcollet},\ and\ \citenamefont {Sachdev}}]{RevModPhys.94.035004}%
  \BibitemOpen
  \bibfield  {author} {\bibinfo {author} {\bibfnamefont {D.}~\bibnamefont {Chowdhury}}, \bibinfo {author} {\bibfnamefont {A.}~\bibnamefont {Georges}}, \bibinfo {author} {\bibfnamefont {O.}~\bibnamefont {Parcollet}}, \ and\ \bibinfo {author} {\bibfnamefont {S.}~\bibnamefont {Sachdev}},\ }\href {\doibase 10.1103/RevModPhys.94.035004} {\bibfield  {journal} {\bibinfo  {journal} {Rev. Mod. Phys.}\ }\textbf {\bibinfo {volume} {94}},\ \bibinfo {pages} {035004} (\bibinfo {year} {2022})}\BibitemShut {NoStop}%
\bibitem [{phi()}]{phi4}%
  \BibitemOpen
  \href@noop {} {}\bibinfo {note} {There is an implicit assumption that high-order magnon-mangon interactions is much smaller than electron-spin exchange interactions due to the renormalization of the effective mass of higher-order magnon-magnon interactions is different from that at zero temperature. However, at the mean-field level, we do not require this assumption.}\BibitemShut {Stop}%
\bibitem [{\citenamefont {Zhou}\ and\ \citenamefont {Wang}(2022)}]{zhou2022chern}%
  \BibitemOpen
  \bibfield  {author} {\bibinfo {author} {\bibfnamefont {S.}~\bibnamefont {Zhou}}\ and\ \bibinfo {author} {\bibfnamefont {Z.}~\bibnamefont {Wang}},\ }\href {\doibase 10.1038/s41467-022-34832-2} {\bibfield  {journal} {\bibinfo  {journal} {Nature Communications}\ }\textbf {\bibinfo {volume} {13}},\ \bibinfo {pages} {7288} (\bibinfo {year} {2022})}\BibitemShut {NoStop}%
\bibitem [{\citenamefont {Hamidian}\ \emph {et~al.}(2016)\citenamefont {Hamidian}, \citenamefont {Edkins}, \citenamefont {Joo}, \citenamefont {Kostin}, \citenamefont {Eisaki}, \citenamefont {Uchida}, \citenamefont {Lawler}, \citenamefont {Kim}, \citenamefont {Mackenzie}, \citenamefont {Fujita} \emph {et~al.}}]{hamidian2016detection}%
  \BibitemOpen
  \bibfield  {author} {\bibinfo {author} {\bibfnamefont {M.}~\bibnamefont {Hamidian}}, \bibinfo {author} {\bibfnamefont {S.}~\bibnamefont {Edkins}}, \bibinfo {author} {\bibfnamefont {S.~H.}\ \bibnamefont {Joo}}, \bibinfo {author} {\bibfnamefont {A.}~\bibnamefont {Kostin}}, \bibinfo {author} {\bibfnamefont {H.}~\bibnamefont {Eisaki}}, \bibinfo {author} {\bibfnamefont {S.}~\bibnamefont {Uchida}}, \bibinfo {author} {\bibfnamefont {M.}~\bibnamefont {Lawler}}, \bibinfo {author} {\bibfnamefont {E.-A.}\ \bibnamefont {Kim}}, \bibinfo {author} {\bibfnamefont {A.}~\bibnamefont {Mackenzie}}, \bibinfo {author} {\bibfnamefont {K.}~\bibnamefont {Fujita}},  \emph {et~al.},\ }\href {\doibase 10.1038/nature17411} {\bibfield  {journal} {\bibinfo  {journal} {Nature}\ }\textbf {\bibinfo {volume} {532}},\ \bibinfo {pages} {343} (\bibinfo {year} {2016})}\BibitemShut {NoStop}%
\bibitem [{\citenamefont {Wang}\ \emph {et~al.}(2015)\citenamefont {Wang}, \citenamefont {Agterberg},\ and\ \citenamefont {Chubukov}}]{PhysRevLett.114.197001}%
  \BibitemOpen
  \bibfield  {author} {\bibinfo {author} {\bibfnamefont {Y.}~\bibnamefont {Wang}}, \bibinfo {author} {\bibfnamefont {D.~F.}\ \bibnamefont {Agterberg}}, \ and\ \bibinfo {author} {\bibfnamefont {A.}~\bibnamefont {Chubukov}},\ }\href {\doibase 10.1103/PhysRevLett.114.197001} {\bibfield  {journal} {\bibinfo  {journal} {Phys. Rev. Lett.}\ }\textbf {\bibinfo {volume} {114}},\ \bibinfo {pages} {197001} (\bibinfo {year} {2015})}\BibitemShut {NoStop}%
\bibitem [{\citenamefont {Webster}\ and\ \citenamefont {Yan}(2018)}]{PhysRevB.98.144411}%
  \BibitemOpen
  \bibfield  {author} {\bibinfo {author} {\bibfnamefont {L.}~\bibnamefont {Webster}}\ and\ \bibinfo {author} {\bibfnamefont {J.-A.}\ \bibnamefont {Yan}},\ }\href {\doibase 10.1103/PhysRevB.98.144411} {\bibfield  {journal} {\bibinfo  {journal} {Phys. Rev. B}\ }\textbf {\bibinfo {volume} {98}},\ \bibinfo {pages} {144411} (\bibinfo {year} {2018})}\BibitemShut {NoStop}%
\bibitem [{\citenamefont {M\ae{}land}\ and\ \citenamefont {Sudb\o{}}(2022)}]{PhysRevResearch.4.L032025}%
  \BibitemOpen
  \bibfield  {author} {\bibinfo {author} {\bibfnamefont {K.}~\bibnamefont {M\ae{}land}}\ and\ \bibinfo {author} {\bibfnamefont {A.}~\bibnamefont {Sudb\o{}}},\ }\href {\doibase 10.1103/PhysRevResearch.4.L032025} {\bibfield  {journal} {\bibinfo  {journal} {Phys. Rev. Res.}\ }\textbf {\bibinfo {volume} {4}},\ \bibinfo {pages} {L032025} (\bibinfo {year} {2022})}\BibitemShut {NoStop}%
\bibitem [{ElE()}]{ElEnSp}%
  \BibitemOpen
  \href@noop {} {}\bibinfo {note} {In fact, there is an additional term $\frac{JS}{\sqrt{2S}}\sum_i(c^\dagger_{i\downarrow}c_{i\downarrow}-c^\dagger_{i\uparrow}c_{i\uparrow})$ in $H_{em}$ that can lead to different energy spectra for electrons with different spins. However, this can be addressed by introducing a magnetic field $B$ to the normal metal layer (with new tuning parameter $\Delta=k_z^2+4K/\bar{J}a^2-2B/S\bar{J}a^2$) or changing the chemical potential. Furthermore, in the subsequent scaling analysis, it can be proven that the NFL/MFL behavior is not sensitive to the specific form of the energy spectrum (, it only depends on the power-law behavior of momentum after a Taylor expansion). So, even when considering scattering between patches near VHPS of Fermi surface with different spin, it does not affect our conclusions.}\BibitemShut {Stop}%
\bibitem [{EP()}]{EP}%
  \BibitemOpen
  \href@noop {} {}\bibinfo {note} {Unless the Fermi surface is nested, in which case the contribution of flat Fermi surface is also important \cite{PhysRevB.64.165107}.}\BibitemShut {Stop}%
\bibitem [{EnE()}]{EnEl}%
  \BibitemOpen
  \href@noop {} {}\bibinfo {note} {For a square lattice considering only nearest-neighbor hopping, the VHPs occur at half-filling, and the electron dispersions around VHPs can be written as $\pm t(k_x^2-k_y^2)$. Dispersion near VHP in Ref. \cite{PhysRevLett.130.126001} is $\tilde{t}(k_x'^2-3k_y'^2)/2$.}\BibitemShut {Stop}%
\bibitem [{\citenamefont {Isobe}\ \emph {et~al.}(2018)\citenamefont {Isobe}, \citenamefont {Yuan},\ and\ \citenamefont {Fu}}]{PhysRevX.8.041041}%
  \BibitemOpen
  \bibfield  {author} {\bibinfo {author} {\bibfnamefont {H.}~\bibnamefont {Isobe}}, \bibinfo {author} {\bibfnamefont {N.~F.~Q.}\ \bibnamefont {Yuan}}, \ and\ \bibinfo {author} {\bibfnamefont {L.}~\bibnamefont {Fu}},\ }\href {\doibase 10.1103/PhysRevX.8.041041} {\bibfield  {journal} {\bibinfo  {journal} {Phys. Rev. X}\ }\textbf {\bibinfo {volume} {8}},\ \bibinfo {pages} {041041} (\bibinfo {year} {2018})}\BibitemShut {NoStop}%
\bibitem [{\citenamefont {Yao}\ and\ \citenamefont {Yang}(2015)}]{PhysRevB.92.035132}%
  \BibitemOpen
  \bibfield  {author} {\bibinfo {author} {\bibfnamefont {H.}~\bibnamefont {Yao}}\ and\ \bibinfo {author} {\bibfnamefont {F.}~\bibnamefont {Yang}},\ }\href {\doibase 10.1103/PhysRevB.92.035132} {\bibfield  {journal} {\bibinfo  {journal} {Phys. Rev. B}\ }\textbf {\bibinfo {volume} {92}},\ \bibinfo {pages} {035132} (\bibinfo {year} {2015})}\BibitemShut {NoStop}%
\bibitem [{SM()}]{SM}%
  \BibitemOpen
  \href@noop {} {}\bibinfo {note} {See Supplemental Material at XXX for the derivation of the self-energies of fermion and boson with translationally invariant SYK-liked Yukawa interaction (S1.1), disordered SYK-liked Yukawa interaction (S1.2) and both (S1.3). The transport properties are shown in (S2). Supplementary information on the phase diagram is presented in (S3).}\BibitemShut {Stop}%
\bibitem [{\citenamefont {Affleck}\ and\ \citenamefont {Ludwig}(1993)}]{PhysRevB.48.7297}%
  \BibitemOpen
  \bibfield  {author} {\bibinfo {author} {\bibfnamefont {I.}~\bibnamefont {Affleck}}\ and\ \bibinfo {author} {\bibfnamefont {A.~W.~W.}\ \bibnamefont {Ludwig}},\ }\href {\doibase 10.1103/PhysRevB.48.7297} {\bibfield  {journal} {\bibinfo  {journal} {Phys. Rev. B}\ }\textbf {\bibinfo {volume} {48}},\ \bibinfo {pages} {7297} (\bibinfo {year} {1993})}\BibitemShut {NoStop}%
\bibitem [{\citenamefont {Gerlach}\ \emph {et~al.}(2017)\citenamefont {Gerlach}, \citenamefont {Schattner}, \citenamefont {Berg},\ and\ \citenamefont {Trebst}}]{PhysRevB.95.035124}%
  \BibitemOpen
  \bibfield  {author} {\bibinfo {author} {\bibfnamefont {M.~H.}\ \bibnamefont {Gerlach}}, \bibinfo {author} {\bibfnamefont {Y.}~\bibnamefont {Schattner}}, \bibinfo {author} {\bibfnamefont {E.}~\bibnamefont {Berg}}, \ and\ \bibinfo {author} {\bibfnamefont {S.}~\bibnamefont {Trebst}},\ }\href {\doibase 10.1103/PhysRevB.95.035124} {\bibfield  {journal} {\bibinfo  {journal} {Phys. Rev. B}\ }\textbf {\bibinfo {volume} {95}},\ \bibinfo {pages} {035124} (\bibinfo {year} {2017})}\BibitemShut {NoStop}%
\bibitem [{\citenamefont {Driskell}\ \emph {et~al.}(2021)\citenamefont {Driskell}, \citenamefont {Lederer}, \citenamefont {Bauer}, \citenamefont {Trebst},\ and\ \citenamefont {Kim}}]{PhysRevLett.127.046601}%
  \BibitemOpen
  \bibfield  {author} {\bibinfo {author} {\bibfnamefont {G.}~\bibnamefont {Driskell}}, \bibinfo {author} {\bibfnamefont {S.}~\bibnamefont {Lederer}}, \bibinfo {author} {\bibfnamefont {C.}~\bibnamefont {Bauer}}, \bibinfo {author} {\bibfnamefont {S.}~\bibnamefont {Trebst}}, \ and\ \bibinfo {author} {\bibfnamefont {E.-A.}\ \bibnamefont {Kim}},\ }\href {\doibase 10.1103/PhysRevLett.127.046601} {\bibfield  {journal} {\bibinfo  {journal} {Phys. Rev. Lett.}\ }\textbf {\bibinfo {volume} {127}},\ \bibinfo {pages} {046601} (\bibinfo {year} {2021})}\BibitemShut {NoStop}%
\bibitem [{\citenamefont {Hertz}(1976)}]{PhysRevB.14.1165}%
  \BibitemOpen
  \bibfield  {author} {\bibinfo {author} {\bibfnamefont {J.~A.}\ \bibnamefont {Hertz}},\ }\href {\doibase 10.1103/PhysRevB.14.1165} {\bibfield  {journal} {\bibinfo  {journal} {Phys. Rev. B}\ }\textbf {\bibinfo {volume} {14}},\ \bibinfo {pages} {1165} (\bibinfo {year} {1976})}\BibitemShut {NoStop}%
\bibitem [{\citenamefont {Lederer}\ \emph {et~al.}(2017)\citenamefont {Lederer}, \citenamefont {Schattner}, \citenamefont {Berg},\ and\ \citenamefont {Kivelson}}]{doi:10.1073/pnas.1620651114}%
  \BibitemOpen
  \bibfield  {author} {\bibinfo {author} {\bibfnamefont {S.}~\bibnamefont {Lederer}}, \bibinfo {author} {\bibfnamefont {Y.}~\bibnamefont {Schattner}}, \bibinfo {author} {\bibfnamefont {E.}~\bibnamefont {Berg}}, \ and\ \bibinfo {author} {\bibfnamefont {S.~A.}\ \bibnamefont {Kivelson}},\ }\href {\doibase 10.1073/pnas.1620651114} {\bibfield  {journal} {\bibinfo  {journal} {Proceedings of the National Academy of Sciences}\ }\textbf {\bibinfo {volume} {114}},\ \bibinfo {pages} {4905} (\bibinfo {year} {2017})}\BibitemShut {NoStop}%
\bibitem [{\citenamefont {Lederer}\ \emph {et~al.}(2015)\citenamefont {Lederer}, \citenamefont {Schattner}, \citenamefont {Berg},\ and\ \citenamefont {Kivelson}}]{PhysRevLett.114.097001}%
  \BibitemOpen
  \bibfield  {author} {\bibinfo {author} {\bibfnamefont {S.}~\bibnamefont {Lederer}}, \bibinfo {author} {\bibfnamefont {Y.}~\bibnamefont {Schattner}}, \bibinfo {author} {\bibfnamefont {E.}~\bibnamefont {Berg}}, \ and\ \bibinfo {author} {\bibfnamefont {S.~A.}\ \bibnamefont {Kivelson}},\ }\href {\doibase 10.1103/PhysRevLett.114.097001} {\bibfield  {journal} {\bibinfo  {journal} {Phys. Rev. Lett.}\ }\textbf {\bibinfo {volume} {114}},\ \bibinfo {pages} {097001} (\bibinfo {year} {2015})}\BibitemShut {NoStop}%
\bibitem [{\citenamefont {Radovan}\ \emph {et~al.}(2003)\citenamefont {Radovan}, \citenamefont {Fortune}, \citenamefont {Murphy}, \citenamefont {Hannahs}, \citenamefont {Palm}, \citenamefont {Tozer},\ and\ \citenamefont {Hall}}]{radovan2003magnetic}%
  \BibitemOpen
  \bibfield  {author} {\bibinfo {author} {\bibfnamefont {H.}~\bibnamefont {Radovan}}, \bibinfo {author} {\bibfnamefont {N.}~\bibnamefont {Fortune}}, \bibinfo {author} {\bibfnamefont {T.}~\bibnamefont {Murphy}}, \bibinfo {author} {\bibfnamefont {S.}~\bibnamefont {Hannahs}}, \bibinfo {author} {\bibfnamefont {E.}~\bibnamefont {Palm}}, \bibinfo {author} {\bibfnamefont {S.}~\bibnamefont {Tozer}}, \ and\ \bibinfo {author} {\bibfnamefont {D.}~\bibnamefont {Hall}},\ }\href {\doibase 10.1038/nature01842} {\bibfield  {journal} {\bibinfo  {journal} {Nature}\ }\textbf {\bibinfo {volume} {425}},\ \bibinfo {pages} {51} (\bibinfo {year} {2003})}\BibitemShut {NoStop}%
\bibitem [{\citenamefont {Agterberg}\ \emph {et~al.}(2020)\citenamefont {Agterberg}, \citenamefont {Davis}, \citenamefont {Edkins}, \citenamefont {Fradkin}, \citenamefont {Van~Harlingen}, \citenamefont {Kivelson}, \citenamefont {Lee}, \citenamefont {Radzihovsky}, \citenamefont {Tranquada},\ and\ \citenamefont {Wang}}]{annurev-conmatphys-031119-050711}%
  \BibitemOpen
  \bibfield  {author} {\bibinfo {author} {\bibfnamefont {D.~F.}\ \bibnamefont {Agterberg}}, \bibinfo {author} {\bibfnamefont {J.~S.}\ \bibnamefont {Davis}}, \bibinfo {author} {\bibfnamefont {S.~D.}\ \bibnamefont {Edkins}}, \bibinfo {author} {\bibfnamefont {E.}~\bibnamefont {Fradkin}}, \bibinfo {author} {\bibfnamefont {D.~J.}\ \bibnamefont {Van~Harlingen}}, \bibinfo {author} {\bibfnamefont {S.~A.}\ \bibnamefont {Kivelson}}, \bibinfo {author} {\bibfnamefont {P.~A.}\ \bibnamefont {Lee}}, \bibinfo {author} {\bibfnamefont {L.}~\bibnamefont {Radzihovsky}}, \bibinfo {author} {\bibfnamefont {J.~M.}\ \bibnamefont {Tranquada}}, \ and\ \bibinfo {author} {\bibfnamefont {Y.}~\bibnamefont {Wang}},\ }\href {\doibase 10.1146/annurev-conmatphys-031119-050711} {\bibfield  {journal} {\bibinfo  {journal} {Annual Review of Condensed Matter Physics}\ }\textbf {\bibinfo {volume} {11}},\ \bibinfo {pages} {231} (\bibinfo {year} {2020})}\BibitemShut {NoStop}%
\bibitem [{\citenamefont {Li}\ \emph {et~al.}(2010)\citenamefont {Li}, \citenamefont {Luican}, \citenamefont {Lopes~dos Santos}, \citenamefont {Castro~Neto}, \citenamefont {Reina}, \citenamefont {Kong},\ and\ \citenamefont {Andrei}}]{li2010observation}%
  \BibitemOpen
  \bibfield  {author} {\bibinfo {author} {\bibfnamefont {G.}~\bibnamefont {Li}}, \bibinfo {author} {\bibfnamefont {A.}~\bibnamefont {Luican}}, \bibinfo {author} {\bibfnamefont {J.}~\bibnamefont {Lopes~dos Santos}}, \bibinfo {author} {\bibfnamefont {A.}~\bibnamefont {Castro~Neto}}, \bibinfo {author} {\bibfnamefont {A.}~\bibnamefont {Reina}}, \bibinfo {author} {\bibfnamefont {J.}~\bibnamefont {Kong}}, \ and\ \bibinfo {author} {\bibfnamefont {E.}~\bibnamefont {Andrei}},\ }\href {\doibase 10.1038/nphys1463} {\bibfield  {journal} {\bibinfo  {journal} {Nature physics}\ }\textbf {\bibinfo {volume} {6}},\ \bibinfo {pages} {109} (\bibinfo {year} {2010})}\BibitemShut {NoStop}%
\bibitem [{\citenamefont {Yan}\ \emph {et~al.}(2012)\citenamefont {Yan}, \citenamefont {Liu}, \citenamefont {Dou}, \citenamefont {Meng}, \citenamefont {Feng}, \citenamefont {Chu}, \citenamefont {Zhang}, \citenamefont {Liu}, \citenamefont {Nie},\ and\ \citenamefont {He}}]{PhysRevLett.109.126801}%
  \BibitemOpen
  \bibfield  {author} {\bibinfo {author} {\bibfnamefont {W.}~\bibnamefont {Yan}}, \bibinfo {author} {\bibfnamefont {M.}~\bibnamefont {Liu}}, \bibinfo {author} {\bibfnamefont {R.-F.}\ \bibnamefont {Dou}}, \bibinfo {author} {\bibfnamefont {L.}~\bibnamefont {Meng}}, \bibinfo {author} {\bibfnamefont {L.}~\bibnamefont {Feng}}, \bibinfo {author} {\bibfnamefont {Z.-D.}\ \bibnamefont {Chu}}, \bibinfo {author} {\bibfnamefont {Y.}~\bibnamefont {Zhang}}, \bibinfo {author} {\bibfnamefont {Z.}~\bibnamefont {Liu}}, \bibinfo {author} {\bibfnamefont {J.-C.}\ \bibnamefont {Nie}}, \ and\ \bibinfo {author} {\bibfnamefont {L.}~\bibnamefont {He}},\ }\href {\doibase 10.1103/PhysRevLett.109.126801} {\bibfield  {journal} {\bibinfo  {journal} {Phys. Rev. Lett.}\ }\textbf {\bibinfo {volume} {109}},\ \bibinfo {pages} {126801} (\bibinfo {year} {2012})}\BibitemShut {NoStop}%
\bibitem [{\citenamefont {Yin}\ \emph {et~al.}(2016)\citenamefont {Yin}, \citenamefont {Wang}, \citenamefont {Peng}, \citenamefont {Tan}, \citenamefont {Liao}, \citenamefont {Lin}, \citenamefont {Sun}, \citenamefont {Koh}, \citenamefont {Chen}, \citenamefont {Peng} \emph {et~al.}}]{yin2016selectively}%
  \BibitemOpen
  \bibfield  {author} {\bibinfo {author} {\bibfnamefont {J.}~\bibnamefont {Yin}}, \bibinfo {author} {\bibfnamefont {H.}~\bibnamefont {Wang}}, \bibinfo {author} {\bibfnamefont {H.}~\bibnamefont {Peng}}, \bibinfo {author} {\bibfnamefont {Z.}~\bibnamefont {Tan}}, \bibinfo {author} {\bibfnamefont {L.}~\bibnamefont {Liao}}, \bibinfo {author} {\bibfnamefont {L.}~\bibnamefont {Lin}}, \bibinfo {author} {\bibfnamefont {X.}~\bibnamefont {Sun}}, \bibinfo {author} {\bibfnamefont {A.~L.}\ \bibnamefont {Koh}}, \bibinfo {author} {\bibfnamefont {Y.}~\bibnamefont {Chen}}, \bibinfo {author} {\bibfnamefont {H.}~\bibnamefont {Peng}},  \emph {et~al.},\ }\href {\doibase 10.1038/ncomms10699} {\bibfield  {journal} {\bibinfo  {journal} {Nature communications}\ }\textbf {\bibinfo {volume} {7}},\ \bibinfo {pages} {10699} (\bibinfo {year} {2016})}\BibitemShut {NoStop}%
\bibitem [{\citenamefont {Xu}\ \emph {et~al.}(2021{\natexlab{b}})\citenamefont {Xu}, \citenamefont {Al~Ezzi}, \citenamefont {Balakrishnan}, \citenamefont {Garcia-Ruiz}, \citenamefont {Tsim}, \citenamefont {Mullan}, \citenamefont {Barrier}, \citenamefont {Xin}, \citenamefont {Piot}, \citenamefont {Taniguchi} \emph {et~al.}}]{xu2021tunable}%
  \BibitemOpen
  \bibfield  {author} {\bibinfo {author} {\bibfnamefont {S.}~\bibnamefont {Xu}}, \bibinfo {author} {\bibfnamefont {M.~M.}\ \bibnamefont {Al~Ezzi}}, \bibinfo {author} {\bibfnamefont {N.}~\bibnamefont {Balakrishnan}}, \bibinfo {author} {\bibfnamefont {A.}~\bibnamefont {Garcia-Ruiz}}, \bibinfo {author} {\bibfnamefont {B.}~\bibnamefont {Tsim}}, \bibinfo {author} {\bibfnamefont {C.}~\bibnamefont {Mullan}}, \bibinfo {author} {\bibfnamefont {J.}~\bibnamefont {Barrier}}, \bibinfo {author} {\bibfnamefont {N.}~\bibnamefont {Xin}}, \bibinfo {author} {\bibfnamefont {B.~A.}\ \bibnamefont {Piot}}, \bibinfo {author} {\bibfnamefont {T.}~\bibnamefont {Taniguchi}},  \emph {et~al.},\ }\href {\doibase 10.1038/s41567-021-01172-9} {\bibfield  {journal} {\bibinfo  {journal} {Nature Physics}\ }\textbf {\bibinfo {volume} {17}},\ \bibinfo {pages} {619} (\bibinfo {year} {2021}{\natexlab{b}})}\BibitemShut {NoStop}%
\bibitem [{\citenamefont {Marsiglio}(2020)}]{MARSIGLIO2020168102}%
  \BibitemOpen
  \bibfield  {author} {\bibinfo {author} {\bibfnamefont {F.}~\bibnamefont {Marsiglio}},\ }\href {\doibase https://doi.org/10.1016/j.aop.2020.168102} {\bibfield  {journal} {\bibinfo  {journal} {Annals of Physics}\ }\textbf {\bibinfo {volume} {417}},\ \bibinfo {pages} {168102} (\bibinfo {year} {2020})},\ \bibinfo {note} {eliashberg theory at 60: Strong-coupling superconductivity and beyond}\BibitemShut {NoStop}%
\bibitem [{\citenamefont {Irkhin}\ \emph {et~al.}(2001)\citenamefont {Irkhin}, \citenamefont {Katanin},\ and\ \citenamefont {Katsnelson}}]{PhysRevB.64.165107}%
  \BibitemOpen
  \bibfield  {author} {\bibinfo {author} {\bibfnamefont {V.~Y.}\ \bibnamefont {Irkhin}}, \bibinfo {author} {\bibfnamefont {A.~A.}\ \bibnamefont {Katanin}}, \ and\ \bibinfo {author} {\bibfnamefont {M.~I.}\ \bibnamefont {Katsnelson}},\ }\href {\doibase 10.1103/PhysRevB.64.165107} {\bibfield  {journal} {\bibinfo  {journal} {Phys. Rev. B}\ }\textbf {\bibinfo {volume} {64}},\ \bibinfo {pages} {165107} (\bibinfo {year} {2001})}\BibitemShut {NoStop}%
\end{thebibliography}%
\clearpage


\makeatletter
\renewcommand{\theequation}{S\arabic{equation}}
\setcounter{equation}{0}
\renewcommand{\thefigure}{S\arabic{figure}}
\setcounter{figure}{0}
\renewcommand{\thesection}{S\arabic{section}}

\onecolumngrid

\section{Self Energy}
  		\subsection{Translationally Invariant SYK-liked Yukawa Interaction}
  		
  Firstly, we consider translationally invariant flavor-random Yukawa interaction:
  \begin{equation}
  	\mathcal{H}_{int1}=-\sum_{mnl}\frac{J_{mnl}}{\sqrt{NN'}}\sum_i c^\dagger_{i,m} c_{i,n} a_{i,l} +h.c.,
  	\label{sic-h}
  \end{equation}
  where $i$ is lattice sites, $m,n=1 . . . N$ are the flavors of fermion field and $l=1...N'$ is the flavors of the scalar field. The space independent coupling $J_{mnl}$ is random in the space of flavors with Gaussian distribution:
  \begin{equation}
  	\overline{J_{mnl}}=0,\ \ \ \  \overline{J_{mnl}J^*_{m'n'l'}}=|J|^2\delta_{mm'}\delta_{nn'}\delta_{ll'}.
  \end{equation}
  
  The boson self energy caused by scattering between patch near the van Hove point is:
  \begin{equation}
  	\Pi_J(i\Omega_m,\vec{q})=-|J|^2\frac{N}{N'}T\sum_n\int\frac{d^2k}{(2\pi)^2}\frac{1}{i\omega_n-k_x^2-ak_y^2}\frac{1}{i\omega_n+i\Omega_m-(k_x+q_x)^2-a(k_y+q_y)^2}.
  \end{equation}
  The scattering between different patches near van Hove points are considered in the manuscript. Sum over Matsubara frequency and takes $T\rightarrow 0$, this gives:
  \begin{equation}
  	|J|^2\frac{N}{N'}\int\frac{d^2k}{(2\pi)^2}\frac{2(q_x^2+a q_y^2-2k_x q_x-2a k_y q_y)}{(q_x^2+a q_y^2-2 k_x q_y-2a k_y q_y)^2+\Omega_m^2},
  \end{equation}
  The integral domain of $d^2k$ is $-\sqrt{-a}|k_y|<k_x<\sqrt{-a}|k_y|$. We first integrate with respect to $k_x$, and then integrate over $k_y$, get:
  \begin{equation}
  	|J|^2\frac{N}{N'}\frac{1}{4\pi^2}[\frac{|\Omega_m|}{|q_x||a q_y-\sqrt{-a}|q_x||}\mathrm{ArcTan}(\frac{q_x^2+a q_y^2}{|\Omega_m|})-\frac{|\Omega_m|}{|q_x||a q_y+\sqrt{-a}|q_x||}\mathrm{ArcTan}(\frac{q_x^2+a q_y^2}{|\Omega_m|})].
  \end{equation}		
In the integral above for $k_y$, we only considered contributions near the Fermi surface and neglect contributions that lead to logarithmic divergences $\sim \mathrm{ln}\Lambda_U$, where $\Lambda_U$ is the momentum cutoff. The contributions far from the Fermi surface can be absorbed by $\Pi(0,0)$, while the remaining part is expressed as powers of $|\Omega_m|^2/(q_x^2+a q_y^2)$ (similar methods can be found in Refs.\cite{PhysRevB.90.161106, PhysRevLett.130.083603}). In the low-frequency limit $|q_x^2+a q_y^2|\gg |\Omega_m|$ and the high-frequency limit $|\Omega_m|\gg |q_x^2+a q_y^2|$, we have, respectively,
 \begin{equation}
 	\begin{matrix}
 |J|^2\frac{N}{N'}\frac{1}{8\pi\sqrt{-a}} \frac{|\Omega_m|}{a q_y^2+q_x^2}(|1-\sqrt{-a}\frac{q_y}{|q_x|}|-|1+\sqrt{-a}\frac{q_y}{|q_x|}|),& \mathrm{when}  \ |q_x^2+a q_y^2|\gg |\Omega_m| \\
 |J|^2\frac{N}{N'}\frac{1}{4\pi^2\sqrt{-a}}\frac{\mathrm{sgn}(q_x^2+a q_y^2)}{|q_x|}(||q_x|-\sqrt{-a}q_y|-||q_x|+\sqrt{-a}q_y|)  & \mathrm{when}  \ |q_x^2+a q_y^2|\ll |\Omega_m|
\end{matrix} .
 \end{equation}	
These limits are all achievable. For example, in the low-frequency limit, although $q_x^2+aq_y^2$ can be zero due to the presence of zero points, as long as the frequency convergence to zero is faster than the momentum approaching zero point, $|\Omega_m|/|q_x^2+a q_y^2|\ll 1$ and holds true.
 
 We only take the low-frequency limit in our calculations. The boson self energy can be expressed in a simpler form based on the division of different regions in the $q_x-q_y$ plane:
 \begin{equation}
 	\Pi_J(i\Omega_m,\vec{q})=\left\{\begin{matrix}
 -|J|^2\frac{N}{N'}\frac{1}{4\pi\sqrt{-a}}\frac{|\Omega_m|}{q_x^2+a q_y^2} & (-\sqrt{-a}q_y<q_x<\sqrt{-a}q_y \ \& \ q_y>0) \equiv \mathbb{A} \\
 |J|^2\frac{N}{N'}\frac{1}{4\pi\sqrt{-a}}\frac{|\Omega_m|}{q_x^2+a q_y^2} & (\sqrt{-a}q_y<q_x<-\sqrt{-a}q_y \ \& \ q_y<0) \equiv \mathbb{B} \\
 -|J|^2\frac{N}{N'}\frac{q_y}{4\pi |q_x|}\frac{|\Omega_m|}{q_x^2+a q_y^2} & (-\frac{|q_x|}{\sqrt{-a}}< q_y<\frac{|q_x|}{\sqrt{-a}}) \equiv \mathbb{C}
\end{matrix}\right.,
\label{sic-bos-1}
 \end{equation}		
  where we label the three different regions as $\mathbb{A}$, $\mathbb{B}$ and $\mathbb{C}$, respectively.	 We argue that non-Fermi-liquid of marginal-Fermi-liquid occur above the critical point of bosons, so fermions couple with critical bosons, i.e., satisfy $\Delta=\Pi(0,0)$ (only consider the Ising order). We define $\tilde{J}^2\equiv |J|^2\frac{N}{N'}\frac{1}{4\pi \sqrt{-a}}$ as simplify. 
  
  The fermion self energy can be written as:
 \begin{equation}
 	\begin{aligned}
 		\Sigma_J(i\omega_n,0)=&|J|^2T\sum_m[\int_\mathbb{A} \frac{d^2 q}{(2\pi)^2}\frac{1}{q^2+\tilde{J}^2\frac{|\Omega_m|}{q_x^2+a q_y^2}}\frac{1}{i\omega_n+i\Omega_m-q_x^2-a q_y^2}+\int_\mathbb{B}\frac{d^2 q}{(2\pi)^2}\frac{1}{q^2-\tilde{J}^2\frac{|\Omega_m|}{q_x^2+a q_y^2}}\frac{1}{i\omega_n+i\Omega_m-q_x^2-a q_y^2}\\
 		&+\int_\mathbb{C}\frac{d^2 q}{(2\pi)^2}\frac{1}{q^2+\tilde{J}^2\frac{\sqrt{-a}q_y}{|q_x|}\frac{|\Omega_m|}{q_x^2+a q_y^2}}\frac{1}{i\omega_n+i\Omega_m-q_x^2-a q_y^2}].
 	\end{aligned}
 \end{equation}
  	
We will provide the steps for integrating over regions $\mathbb{A}$ and $\mathbb{C}$, and the integration over region $\mathbb{B}$, which is similar to region $\mathbb{A}$, will be omitted:
\begin{equation}
		\int_\mathbb{A} \frac{d^2 q}{(2\pi)^2}\frac{1}{q^2+\tilde{J}^2\frac{|\Omega_m|}{q_x^2+a q_y^2}}\frac{1}{i\omega_n+i\Omega_m-q_x^2-a q_y^2}
		\approx \frac{1}{2\pi^2}\int_0^\infty d q_y\frac{1}{q_y^2-\tilde{J}^2\frac{|\Omega_m|}{-a q_y^2}}\frac{\mathrm{ArcTanh}(\frac{\sqrt{-a}q_y}{\sqrt{i\omega_n+i\Omega_m-a q_y^2}})}{\sqrt{i\omega_n+i\Omega_m-a q_y^2}},
	\label{sic-fer-4}	
\end{equation} 	
where we neglect the $q_x$ component of the damped boson propagator in region $\mathbb{A}$, because the main contribution of the integral is concentrated near the positive $q_y$-axis. When the integration region approaches the boundary of the $\mathcal{A}$, the denominator of the boson propagator diverges. In low-frequency limit $|\omega_n+\Omega_m| \ll -a q_y^2$, we need to define a branch cut on the negative $x$-axis to calculate the square root, $\sqrt{i\omega_n+i\Omega_m-a q_y^2}\approx \sqrt{-a}q_y \mathrm{e}^{i\mathrm{sgn}(\omega_n+\Omega_m)\frac{\eta}{2}}$ with $\eta\equiv |\omega_n+\Omega_m|/(-a)q_y^2\rightarrow  0^+$, and logarithmic function. The exponential function $\mathrm{e}^{i\mathrm{sgn}(\omega_n+\Omega_m)\frac{\eta}{2}}$ is essential because the integral from $0$ to $\infty$ concerning $1/(q_y^3-1/q_y)$ diverges, but transformation $\sqrt{-a}\rightarrow \sqrt{-a}\mathrm{e}^{i\mathrm{sgn}(\omega_n+\Omega_m)\frac{\eta}{2}}$ can be performed to avoid divergence, resulting in:
\begin{equation}
	-\frac{1}{32|\tilde{J}||\Omega_m|^\frac{1}{2}}[1+i\mathrm{sgn}(\omega_n+\Omega_m)\ln \Lambda].
	\label{sic-fer-1}
\end{equation}
The $\ln \Lambda\approx \ln 4(-a)\Lambda_U^2/|\omega_n+\Omega_m|$ at the low-frequency limit, where $\Lambda_{U}$ is the UV cut off of the momentum. Similarly, the integral over the domain $\mathbb{B}$ yields:
\begin{equation}
	-\frac{1}{32|\tilde{J}||\Omega_m|^\frac{1}{2}}[\mathrm{sgn}(\omega_n+\Omega_m)i-\ln \Lambda].
	\label{sic-fer-2}
\end{equation}

The integral over the domain $\mathbb{C}$ (a better approximation is the boson propagator with $q_y^2\ll q_x^2$ in domain $\mathbb{C}$):
\begin{equation}
	\begin{aligned}
	&\int_\mathbb{C}\frac{d^2 q}{(2\pi)^2}\frac{1}{q^2+\tilde{J}^2\frac{\sqrt{-a}q_y}{|q_x|}\frac{|\Omega_m|}{q_x^2+a q_y^2}}\frac{1}{i\omega_n+i\Omega_m-q_x^2-a q_y^2}\\
	\approx &\frac{1}{\pi^2\sqrt{-a}}\int_0^\infty d q_x\frac{\frac{1}{\sqrt{i\omega_n+i\Omega_m-q_x^2}}\mathrm{ArcTan}(\frac{q_x}{\sqrt{i\omega_n+i\Omega_m-q_x^2}})+\frac{\tilde{J}^2|\Omega_m|}{q_x^5}\mathrm{ArcTanh}(\frac{\tilde{J}^2|\Omega_m|}{q_x^4})}{q_x^2+\frac{\tilde{J}^4|\Omega_m|^2}{q_x^8}(i\omega_n+i\Omega_m-q_x^2)}\\
	\approx & \frac{1}{32\sqrt{-a}|\tilde{J}||\Omega_m|^{\frac{1}{2}}}[-1+i\mathrm{sgn}(\omega_n+\Omega_m)\ln \Lambda]-\frac{1}{32\sqrt{-a}|\tilde{J}||\Omega_m|^{\frac{1}{2}}}[\mathrm{sgn}(\omega_n+\Omega_m)i+\ln \Lambda]-\frac{1}{2\pi^2\sqrt{-a}\Lambda_I},
	\label{sic-fer-3}
	\end{aligned}
\end{equation}
where we use the approximation $\tilde{J}|\Omega_m|^{1/2}\ll q_x^2$ when dealing with $\mathrm{ArcTanh}(\tilde{J}^2|\Omega_m|/q_x^4)$, which leads to an infrared divergence. We choose an infrared cutoff $\Lambda_I\gg \tilde{J}|\Omega_m|^{1/2} $ and it makes no contribution to the imaginary part final result after the frequency integration and, under the minimal subtraction renormalization scheme, has no contribution to the real part either. Another way to avoid the infrared divergence is $\tilde{J}|\Omega_m|^{1/2}\gg q_x^2$, which, after frequency integration, yields the same imaginary part of the fermion self-energy.

At zero temperature, the integral over frequencies yields:
\begin{equation}
	\Sigma_J(i\omega_n,0)-\Sigma_J(0,0)\approx -i\frac{ |J|\ln\Lambda}{8\sqrt{\pi}}\sqrt{\frac{N'}{N}}[(-a)^{1/4}-(-a)^{-1/4}]\mathrm{sgn}(\omega_n)|\omega_n|^{1/2}-i\frac{|J|}{8\sqrt{\pi}}\sqrt{\frac{N'}{N}}[(-a)^{1/4}+(-a)^{-1/4}]\mathrm{sgn}(\omega_n)|\omega_n|^{1/2},
		\label{sic-fer-5}
 \end{equation}
where we use the minimal subtraction renormalization scheme to eliminate the divergence in the real part. When $a=-1$, Eq. (\ref{sic-fer-5}) becomes
\begin{equation}
	\Sigma_J(i\omega_n,0)-\Sigma_J(0,0)\approx - i\frac{|J|}{8\sqrt{\pi}}\sqrt{\frac{N'}{N}}\mathrm{sgn}(\omega_n)|\omega_n|^{1/2},
	\label{sic-fer-4}
\end{equation}
the divergence $\ln \Lambda$ has been offset.

Anyway, we obtained $\mathrm{Im}\Sigma(i\omega_n,0)\sim |\omega_n|^{1/2}$, just as in the scaling analysis.

\subsection{Disordered SYK-liked Yukawa Interaction}  		
The disordered flavor-random Yukawa interaction is
\begin{equation}
	  	\mathcal{H}_{int2}=-\frac{1}{\sqrt{NN'}}\sum_{mnl}\sum_i J'_{i,mnl} c^\dagger_{i,m} c_{i,n} a_{i,l} +h.c.,
	  	\label{sdc-h}
\end{equation}
where $i$ is lattice sites, $m,n=1 . . . N$ are the flavors of fermion field and $l=1...N'$ is the flavors of the scalar field. The space dependent coupling $J'_{i,mnl}$ is random in the space of flavors with Gaussian distribution:
\begin{equation}
	\overline{J'_{i,mnl}=0}, \ \ \ \ \overline{J'_{i,mnl}J'_{i',m'n'l'}}=|J'|^2\delta_{ii'}\delta_{mm'}\delta_{nn'}\delta_{ll'}.
\end{equation}
The boson self energy caused by scattering between patch near the van Hove point is:
\begin{equation}
	\begin{aligned}
		\Pi_{J'}(i\Omega_m)-\Pi_{J'}(0)&=-|J'|^2\frac{N}{N'}T\sum_n \int \frac{d^2q}{(2\pi)^2}\frac{1}{i\omega_n-q_x^2-a q_y^2}\int \frac{d^2k}{(2\pi)^2}[\frac{1}{i\omega_n+i\Omega_m-k_x^2-a k_y^2}-\frac{1}{i\omega_n-k_x^2-a k_y^2}]\\
		&\approx -\frac{|J'|^2(\ln \Lambda)^2N}{(-a)\pi^3N'}|\Omega_m|,
	\end{aligned}
	\label{sdc-bos}
\end{equation}
where we have used an approach $\ln \Lambda\approx \ln 4(-a)\Lambda_{U}^2/|\omega_n|$ as a constant with the momentum UV cutoff $\Lambda_{U}$ at the low-frequency limit. The numerical calculation yields $\ln \Lambda\approx \ln 8(-a)\Lambda_{U}^2/|\Omega_m|$, and at the linear order level, we take it as a truncation constant. The critical fluctuations of bosonic degrees of freedom are used to eliminate electronic quasiparticles, so non-Fermi-liquid/marginal-Fermi-liquid behavior occurs near the phase transition critical point $\Pi_{J'}(0)=\Delta$ at zero temperature. We define $\tilde{J}'^2\equiv |J'|^2N(\ln \Lambda)^2/(-a)\pi^3 N'$ as simplify. 

The fermion self energy can be written as:
\begin{equation}
	\Sigma_{J'}(i\omega_n)-\Sigma_{J'}(0)=|J'|^2T\sum_m \int \frac{d^2 k}{(2\pi)^2}\frac{1}{k^2+\tilde{J}'^2|\Omega_m|}\int \frac{d^2q}{(2\pi)^2}[\frac{1}{i\omega_n+i\Omega_m-q_x^2-a q_y^2}-\frac{1}{i\Omega_m-q_x^2-a q_y^2}],
		\label{sdc-fer-1}
\end{equation}
where we use the minimal subtraction renormalization scheme to eliminate the frequency divergence in the real part and only obtain the imaginary part of the Fermi self-energy:
\begin{equation}
	\Sigma_{J'}(i\omega_n)-\Sigma_{J'}(0)=-i\frac{|J'|^2\ln\Lambda}{4\pi^3\sqrt{-a}}\omega_n\ln(\frac{e\Lambda_U^2}{\tilde{J}'^2|\omega_n|}),
	\label{sdc-fer-2}
\end{equation}
where $\Lambda_U$ is the momentum UV cutoff for the first integral in Eq. (\ref{sdc-fer-1}).

\subsection{Both translationally invariant and disordered SYK-liked Yukawa coupling}
If we consider both translationally invariant and disordered flavor-random Yukawa coupling (\ref{sic-h}) and (\ref{sdc-h}). The boson self energy can be obtained from previous results:
\begin{equation}
	\Pi(i\Omega_m,\vec{q})-\Pi(0,0)=  |J|^2\frac{N}{N'}\frac{1}{8\pi\sqrt{-a}} \frac{|\Omega_m|}{a q_y^2+q_x^2}(|1-\sqrt{-a}\frac{q_y}{|q_x|}|-|1+\sqrt{-a}\frac{q_y}{|q_x|}|)-\tilde{J}'^2|\Omega_m|.
\end{equation}
Then the fermion self energy can be written as $\Sigma(i\omega_n,0)=\Sigma_1(i\omega_n,0)+\Sigma_2(i\omega_n)$ with
\begin{equation}
	\begin{aligned}
 		\Sigma_1(i\omega_n,0)=&|J|^2T\sum_m[\int_\mathbb{A} \frac{d^2 q}{(2\pi)^2}\frac{1}{q^2+\tilde{J}^2\frac{|\Omega_m|}{q_x^2+a q_y^2}+\tilde{J}'^2|\Omega_m|}\frac{1}{i\omega_n+i\Omega_m-q_x^2-a q_y^2}\\
 		&+\int_\mathbb{B}\frac{d^2 q}{(2\pi)^2}\frac{1}{q^2-\tilde{J}^2\frac{|\Omega_m|}{q_x^2+a q_y^2}+\tilde{J}'^2|\Omega_m|}\frac{1}{i\omega_n+i\Omega_m-q_x^2-a q_y^2}\\
 		&+\int_\mathbb{C}\frac{d^2 q}{(2\pi)^2}\frac{1}{q^2+\tilde{J}^2\frac{\sqrt{-a}q_y}{|q_x|}\frac{|\Omega_m|}{q_x^2+a q_y^2}+\tilde{J}'^2|\Omega_m|}\frac{1}{i\omega_n+i\Omega_m-q_x^2-a q_y^2}],
 	\end{aligned}
 	\label{bo-fer1}
\end{equation}
and
\begin{equation}
	\begin{aligned}
		\Sigma_2(i\omega_n,0)=&|J'|^2T\sum_m[\int_\mathbb{A} \frac{d^2 k}{(2\pi)^2}\frac{1}{k^2+\tilde{J}^2\frac{|\Omega_m|}{k_x^2+a k_y^2}+\tilde{J}'^2|\Omega_m|}+\int_\mathbb{B}\frac{d^2 k}{(2\pi)^2}\frac{1}{k^2-\tilde{J}^2\frac{|\Omega_m|}{k_x^2+a k_y^2}+\tilde{J}'^2|\Omega_m|}\\
 		&+\int_\mathbb{C}\frac{d^2 k}{(2\pi)^2}\frac{1}{k^2+\tilde{J}^2\frac{\sqrt{-a}k_y}{|k_x|}\frac{|\Omega_m|}{k_x^2+a k_y^2}+\tilde{J}'^2|\Omega_m|}]\int \frac{d^2q}{(2\pi)^2}\frac{1}{i\omega_n+i\Omega_m-q_x^2-a q_y^2}.	
	\end{aligned}
	\label{bo-fer2}
\end{equation}

Similar to the treatment in the section 1.1, we can obtain the integrals for each region:
\begin{equation}
	\begin{aligned}
		&\int_\mathbb{A} \frac{d^2 q}{(2\pi)^2}\frac{1}{q^2+\tilde{J}^2\frac{|\Omega_m|}{q_x^2+a q_y^2}+\tilde{J}'^2|\Omega_m|}\frac{1}{i\omega_n+i\Omega_m-q_x^2-a q_y^2}+\int_\mathbb{B}\frac{d^2 q}{(2\pi)^2}\frac{1}{q^2-\tilde{J}^2\frac{|\Omega_m|}{q_x^2+a q_y^2}+\tilde{J}'^2|\Omega_m|}\frac{1}{i\omega_n+i\Omega_m-q_x^2-a q_y^2}\\
		\approx & \frac{1}{\pi^2}\int_0^\infty d q_y\frac{1}{q_y^2+\tilde{J}'^2|\Omega_m|-\frac{\tilde{J}^4|\Omega_m|^2}{a^2 q_y^4(q_y^2+\tilde{J}'^2|\Omega_m|)}}\frac{\mathrm{ArcTanh}(\frac{\sqrt{-a}q_y}{\sqrt{i\omega_n+i\Omega_m-a q_y^2}})}{\sqrt{i\omega_n+i\Omega_m-a q_y^2}}\\
		\approx & \left\{\begin{matrix}
 					\frac{1}{32|\tilde{J}||\Omega_m|^{\frac{1}{2}}}[-1-i\mathrm{sgn}(\omega_n+\Omega_m)][1+i\mathrm{sgn}(\omega_n+\Omega_m)\ln \Lambda] &  \ \ \ \tilde{J}^2\gg (-a\tilde{J}'^4|\Omega_m|) \\
					\frac{1}{16\sqrt{-a}\tilde{J}'^2|\Omega_m|}[-1-i\frac{2\mathrm{sgn}(\omega_n+\Omega_m)}{\pi}\ln \frac{\tilde{J}^2}{(-a)\tilde{J}'^4|\Omega_m|}][1+i\mathrm{sgn}(\omega_n+\Omega_m)\ln \Lambda ]&\ \ \ \tilde{J}^2\ll (-a\tilde{J}'^4|\Omega_m|).
				\end{matrix}\right.				
	\end{aligned}
	\label{bo-fer1-1}
\end{equation}
The same low-frequency condition has been applied above and the integral over the domain $\mathbb{C}$ is 
\begin{equation}
	\begin{aligned}
		&\int_\mathbb{C}\frac{d^2 q}{(2\pi)^2}\frac{1}{q^2+\tilde{J}^2\frac{\sqrt{-a}q_y}{|q_x|}\frac{|\Omega_m|}{q_x^2+a q_y^2}+\tilde{J}'^2|\Omega_m|}\frac{1}{i\omega_n+i\Omega_m-q_x^2-a q_y^2}\\
		\approx &\frac{1}{\pi^2\sqrt{-a}}\int_0^\infty d q_x\frac{\frac{1}{\sqrt{i\omega_n+i\Omega_m-q_x^2}}\mathrm{ArcTan}(\frac{q_x}{\sqrt{i\omega_n+i\Omega_m-q_x^2}})+\frac{\tilde{J}^2|\Omega_m|}{q_x^3(q_x^2+\tilde{J}'^2|\Omega_m|)}\mathrm{ArcTanh}(\frac{\tilde{J}^2|\Omega_m|}{q_x^2(q_x^2+\tilde{J}'^2|\Omega_m|)})}{q_x^2+\tilde{J}'^2|\Omega_m|+\frac{\tilde{J}^4|\Omega_m|^2}{q_x^6(q_x^2+\tilde{J}'^2|\Omega_m|)}(i\omega_n+i\Omega_m-q_x^2)}\\
		\approx & \left\{\begin{matrix}
 					\frac{1}{32\sqrt{-a}|\tilde{J}||\Omega_m|^{\frac{1}{2}}}[-1-i\mathrm{sgn}(\omega_n+\Omega_m)][1-i\mathrm{sgn}(\omega_n+\Omega_m)\ln \Lambda] &  \ \ \ \tilde{J}^2\gg \tilde{J}'^4|\Omega_m| \\
					\frac{1}{16\sqrt{-a}\tilde{J}'^2|\Omega_m|}[-1-i\frac{2\mathrm{sgn}(\omega_n+\Omega_m)}{\pi}\ln \frac{\tilde{J}^2}{\tilde{J}'^4|\Omega_m|}][1-i\mathrm{sgn}(\omega_n+\Omega_m)\ln \Lambda ]&\ \ \ \tilde{J}^2\ll \tilde{J}'^4|\Omega_m|.
				\end{matrix}\right.
	\end{aligned}
	\label{bo-fer1-2}
\end{equation}
Eqs. (\ref{bo-fer1-1}) and (\ref{bo-fer1-2}) have almost the same form with the exception of the sign in front of the $\ln \Lambda$ and it can be exact cancellation when $a=-1$. Integrals over the domain $\mathbb{A}$ and $\mathbb{B}$ in Eq. (\ref{bo-fer2}) is

	\begin{figure}
		\centering
		\includegraphics[width=0.5\textwidth]{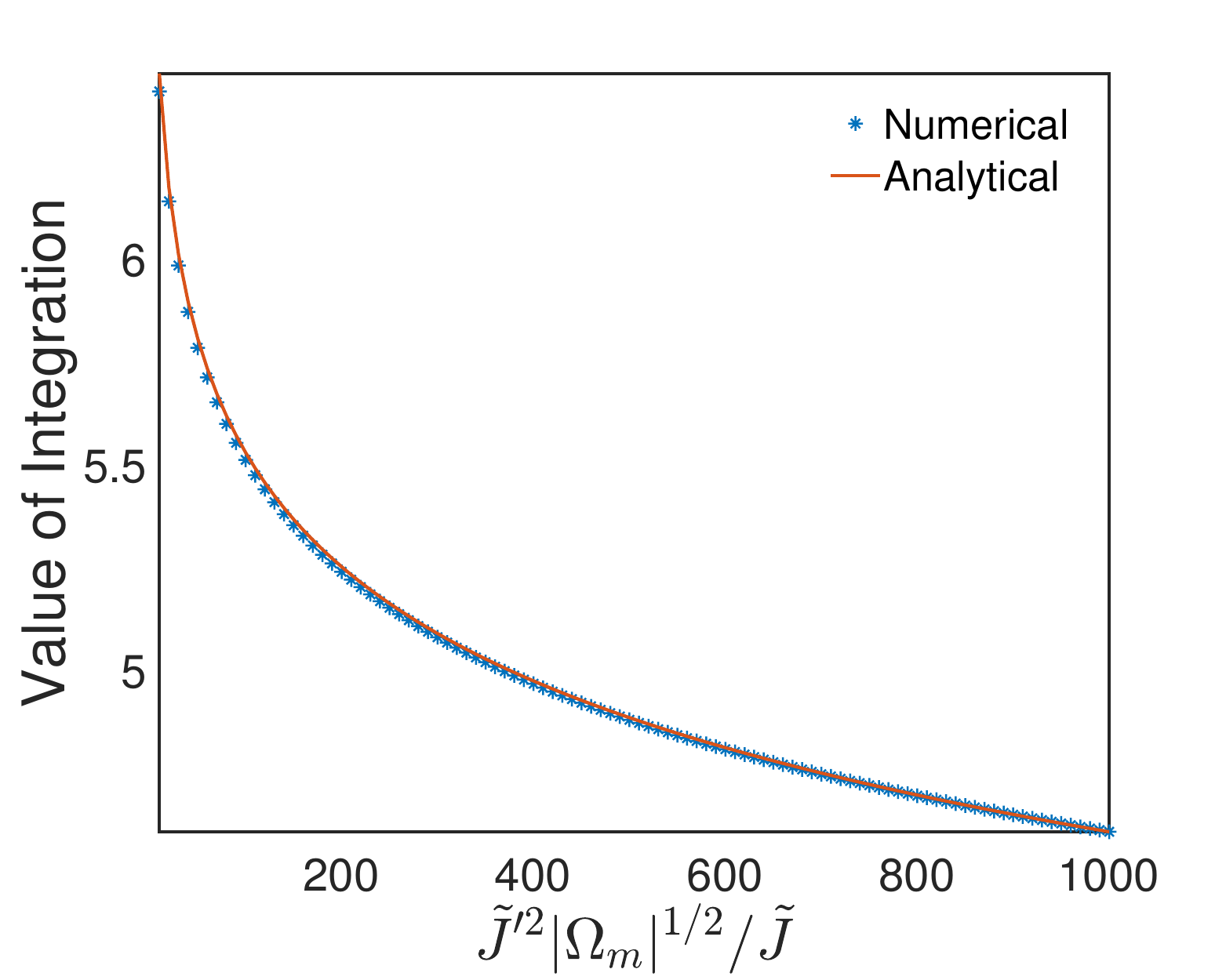}\includegraphics[width=0.5\textwidth]{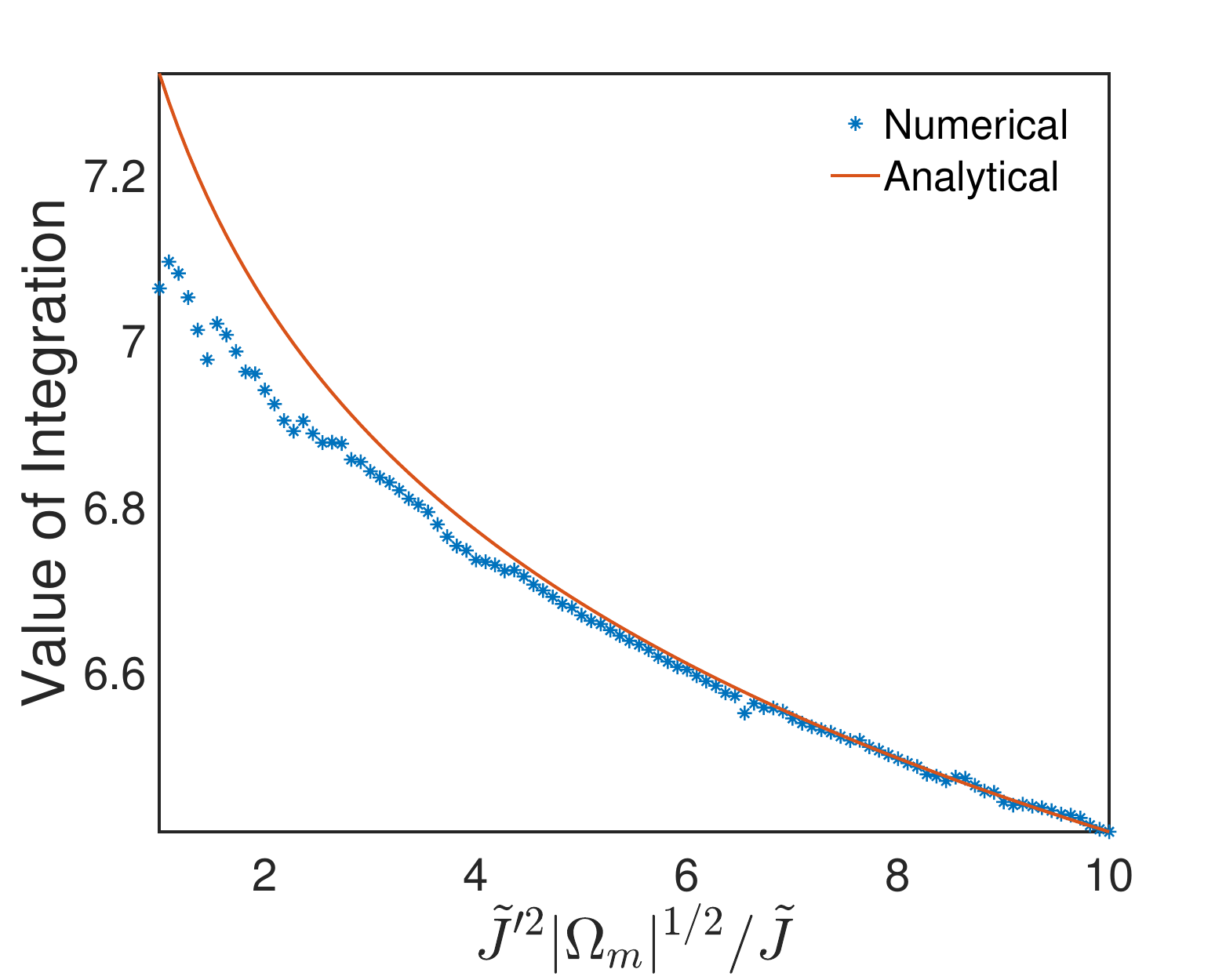}	
		\includegraphics[width=0.5\textwidth]{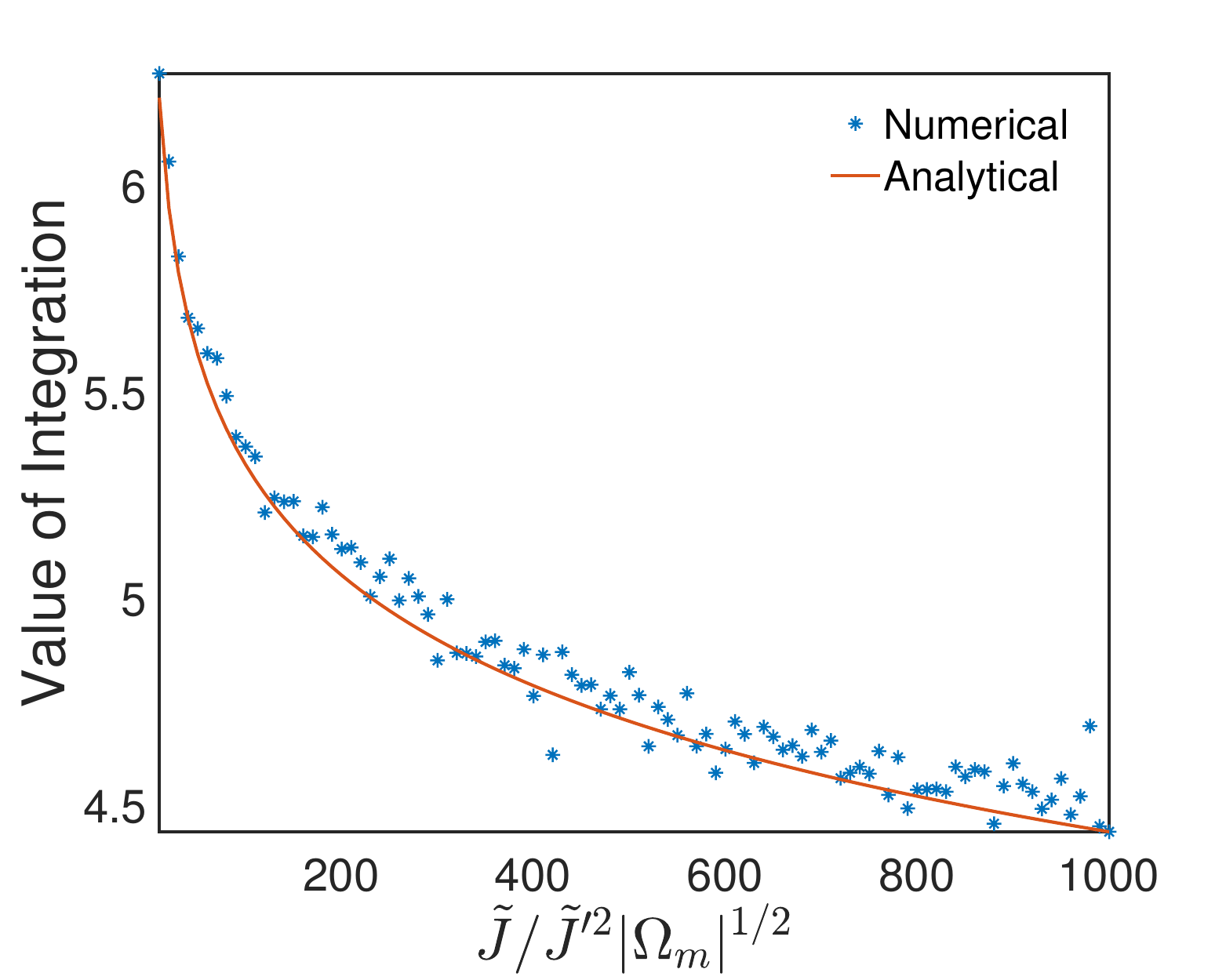}\includegraphics[width=0.5\textwidth]{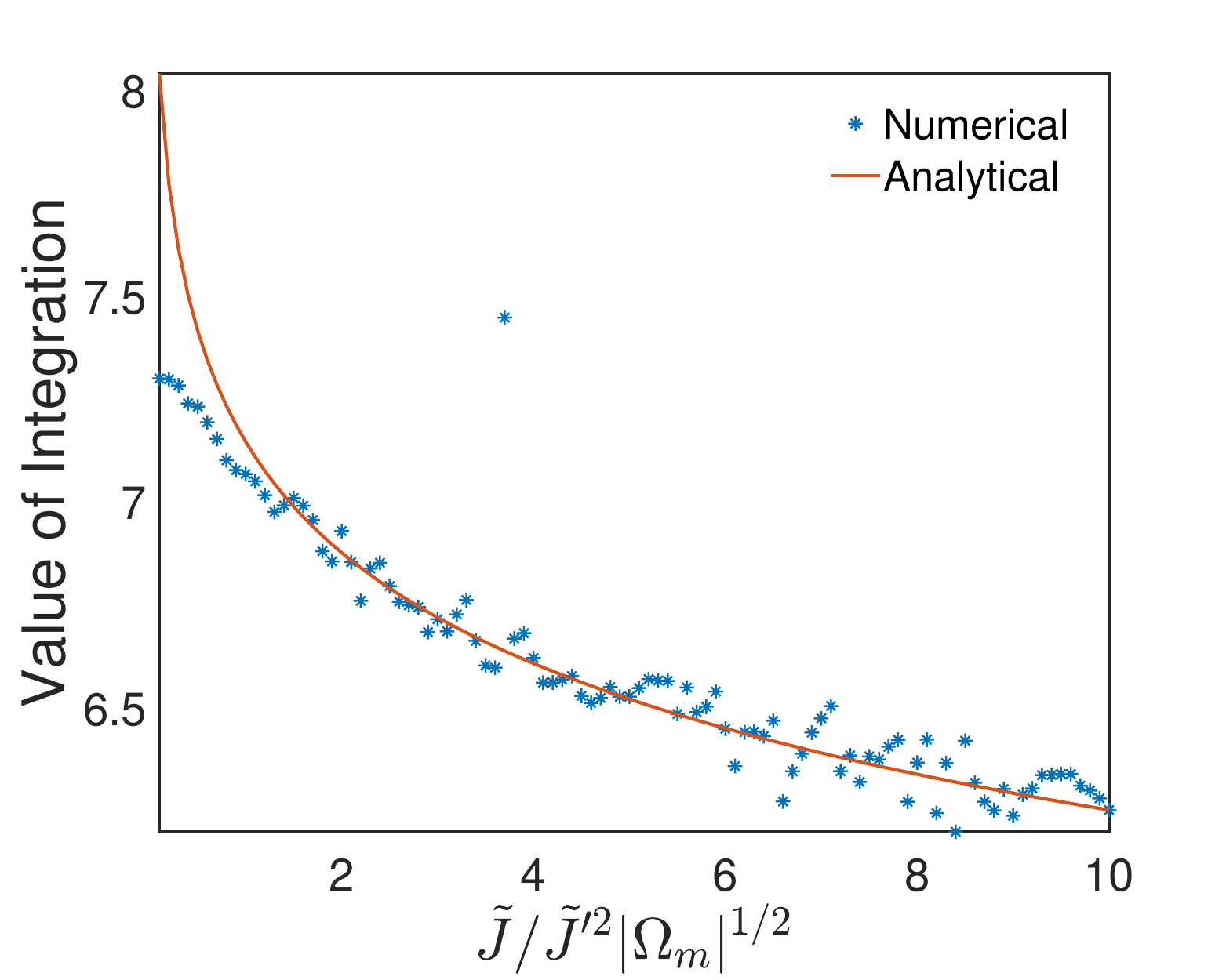}	
		\caption{Comparison between the approximate analytical results and numerical results of the integrals for different limits in Eq. (\ref{bo-fer2-1}). The top two graphs are the results for $\tilde{J}'^2|\Omega_m|^{1/2}\gg \tilde{J}$, with parameters set to $a=-1$, $\Lambda_U/\tilde{J}^{1/2}|\Omega_m|^{1/4}=10^4$. The graph on the top right tells us that the analytical approximation fails at $\tilde{J}'^2|\Omega_m|^{1/2}<5 \tilde{J}$. The bottom two graphs are the results for $\tilde{J}'^2|\Omega_m|^{1/2}\ll \tilde{J}$, with parameters set to $a=-1$, $\Lambda_U/\tilde{J}'|\Omega_m|^{1/2}=10^4$. The graph on the top right tells us that the analytical approximation fails at $\tilde{J}'^2|\Omega_m|^{1/2}>0.5 \tilde{J}$. (When $\tilde{J}'^2|\Omega_m|^{1/2}\ll \tilde{J}$, the singularity is in the large momentum region, and the instability of the integral at the singularity results in poor accuracy in the two graphs below.) } 
		\label{bo-fer2-fig}
	\end{figure}
	
\begin{equation}
	\begin{aligned}
		& \int_\mathbb{A}  \frac{d^2 k}{(2\pi)^2}\frac{1}{k^2+\tilde{J}^2\frac{|\Omega_m|}{k_x^2+a k_y^2}+\tilde{J}'^2|\Omega_m|}+\int_\mathbb{B}\frac{d^2 k}{(2\pi)^2}\frac{1}{k^2-\tilde{J}^2\frac{|\Omega_m|}{k_x^2+a k_y^2}+\tilde{J}'^2|\Omega_m|}\\
		\approx & \frac{1}{\pi^2}\int_0^\infty dk_y \frac{\mathrm{ArcTan}[\frac{\sqrt{-a}k_y}{k_y^2+\tilde{J}'^2|\Omega_m|-\frac{\tilde{J}^4|\Omega_m|^2}{a^2k_y^4(k_y^2+\tilde{J}'^2|\Omega_m|)}}]}{\sqrt{k_y^2+\tilde{J}'^2|\Omega_m|-\frac{\tilde{J}^4|\Omega_m|^2}{a^2k_y^4(k_y^2+\tilde{J}'^2|\Omega_m|)}}}
		\approx \left\{\begin{matrix}
 				 \frac{\mathrm{ArcTan}(\sqrt{-a})}{2\pi^2} \ln \frac{\Lambda_U^2}{\tilde{J}|\Omega_m|^{1/2}} & \ \ \  \tilde{J}^2\gg (-a\tilde{J}'^4|\Omega_m|)\\
 				\frac{\mathrm{ArcTan}(\sqrt{-a})}{\pi^2} \ln \frac{\Lambda_U}{\tilde{J}'|\Omega_m|^{1/2}} & \ \ \  \tilde{J}^2\ll (-a\tilde{J}'^4|\Omega_m|).
\end{matrix}\right.
	\end{aligned}
	\label{bo-fer2-1}
\end{equation}
The superficial degree of divergence of Eq. (\ref{bo-fer2-1}) is $\ln \Lambda_U$. In order to avoid introducing higher-order divergences, we have omitted the $k_x$ component of the second term in the denominator of the first line of the integral expression and retained only the logarithmic divergence part in the handling of the second expression (where $\Lambda_U$ is the UV cutoff). The comparison between the analytical and numerical results can be seen in Fig. (\ref{bo-fer2-fig}), and it can be observed that they agree well under the conditions of $\tilde{J}'^2|\Omega_m|^{1/2}<5 \tilde{J}$ or $\tilde{J}'^2|\Omega_m|^{1/2}>0.5 \tilde{J}$ when $a=-1$. A similar process applies to the integration over the domain $\mathrm{C}$ in Eq. (\ref{bo-fer2}), and yields
\begin{equation}
	\left\{\begin{matrix}
 				 \frac{\mathrm{ArcTan}(1/\sqrt{-a})}{2\pi^2} \ln \frac{\Lambda_U^2}{\tilde{J}|\Omega_m|^{1/2}} & \ \ \  \tilde{J}^2\gg \tilde{J}'^4|\Omega_m|\\
 				\frac{\mathrm{ArcTan}(1/\sqrt{-a})}{\pi^2} \ln \frac{\Lambda_U}{\tilde{J}'|\Omega_m|^{1/2}} & \ \ \  \tilde{J}^2\ll \tilde{J}'^4|\Omega_m|\\
\end{matrix}\right.
\end{equation}

After summing over Matsubara frequency, the fermion self-energy $\Sigma(i\omega_n,0)-\Sigma(0,0)$ at zero temperature is found to be
\begin{equation}
	\begin{aligned}
		 & -i\frac{ |J|\ln \Lambda }{8\sqrt{\pi}}\sqrt{\frac{N'}{N}}[(-a)^{1/4}-(-a)^{-1/4}]\mathrm{sgn}(\omega_n)|\omega_n|^{1/2}-i\frac{|J|}{8\sqrt{\pi}}\sqrt{\frac{N'}{N}}[(-a)^{1/4}+(-a)^{-1/4}]\mathrm{sgn}(\omega_n)|\omega_n|^{1/2}\\
		&-i\frac{|J'|^2\ln \Lambda}{4\pi^4\sqrt{-a}}[\mathrm{ArcTan}(\sqrt{-a})+	\mathrm{ArcTan}(\frac{1}{\sqrt{-a}})]\omega_n\ln\frac{4e\pi\sqrt{-a}N'\Lambda_U^4}{N|J|^2|\omega_n|}
	\end{aligned}
\end{equation}
when $\tilde{J}^2\gg \max\{\tilde{J}'^4|\Omega_m|,(-a)\tilde{J}'^4|\Omega_m|\}$, and 
\begin{equation}
	\begin{aligned}
		&-i\frac{\sqrt{-a}\pi N'\mathrm{sgn}(\omega_n)}{8(\ln\Lambda)^2N|J'|^2}[\ln(-a)\ln\frac{\pi^5(\sqrt{-a})^3N'|J|^2}{4(\ln\Lambda)^4N|J'|^4|\omega_n|}-(\ln\frac{\pi^5(\sqrt{-a})^3N'|J|^2}{4(\ln\Lambda)^4N|J'|^4|\omega_n|})^2]\\
		&-i\frac{|J'|^2\ln \Lambda}{2\pi^4\sqrt{-a}}[\mathrm{ArcTan}(\sqrt{-a})+	\mathrm{ArcTan}(\frac{1}{\sqrt{-a}})]\omega_n\ln\frac{e\pi^3(-a)N'\Lambda_U^2}{N(\ln \Lambda)^2|J'|^2|\omega_n|}
	\end{aligned}
\end{equation}
when $\tilde{J}^2\ll \min\{\tilde{J}'^4|\Omega_m|,(-a)\tilde{J}'^4|\Omega_m|\}$. Eq. (\ref{bo-fer1}) involves an IR divergence in the frequency integration when $\tilde{J}^2\ll \min\{\tilde{J}'^4|\Omega_m|,(-a)\tilde{J}'^4|\Omega_m|\}$, naturally introducing an infrared cutoff $\max\{\tilde{J}^2/\tilde{J}'^4,\tilde{J}^2/(-a)\tilde{J}'^4\}$. But this treatment is not crucial, as analyzed in the manuscript, because the scaling of $\ln\omega$ is concealed by the scaling of marginal-Fermi-liquid.

\section{Transport properties}
We will calculate the conductivity and frequency dependence at zero temperature and extend these results to finite temperature. At finite temperature, the bosonic modes with a dynamic critical exponent $z=2$ (i.e. $\omega\sim\Delta$), where the magnons' self-energy correction (\ref{sdc-bos}) is following a quadratic dispersion relation, acquire a thermal tuning parameter $\tilde{\Delta}(T)\sim T$ (it can be verified by the temperature $T$-dependent phase boundaries between the $U(1)$ phase and the BEC phase, as illustrated in Fig. 3(b) and 3(c) of the main text). This lead to $\omega/T$ scaling in universal function \cite{science.abq6011,PhysRevB.103.235129}, the frequency dependence $\sim \omega^\alpha$ at the limit $T\ll |\omega|$, will transition to a dependence on the temperature $\sim T^\alpha$ at limit $T\gg |\omega|$.

We need to calculate the current-current correlation function to compute the electrical resistivity. By gauging the global $U(1)$ symmetry to introduce gauge field, we assume that the gauge potential $A_k$ exists only at the midpoint of the lattice links. Taylor expansion up to the linear order in $A_k$, we obtain the current:
\begin{equation}
	\vec{j}_k=2\sum_{p,m} \sin((\vec{p}+\frac{\vec{k}}{2})\cdot\vec{e})c^\dagger_{p+k,m}c_{p,m}\vec{e},
\end{equation}  
where $\vec{e}$ is the lattice vector. We assume the low-momentum scenario and consider the gauge potential's direction to be in the x-direction, then we get $j_k=\sum_{p,m} (2p_x+k_x)c^\dagger_{p+k,m}c_{p,m}$. We can also obtain this electromagnetic current from an effective theory (6) of the main text with minimal coupling method. Similarly, it requires fixing the gauge to $A_y=0$. All subsequent calculations involve only the interaction term $H_{int2}$ (i.e. $J=0$) and are at low-frequency limit.

For scattering between the same patch near one van Hove point, the one-loop diagram as shown in Fig1. (b) is automatically zero, as it involves only forward scattering processes and lacks Umklapp scattering processes:
\begin{equation}
	\Xi_b(i\Omega_m,0)=-NT\sum_n\int \frac{d^2k}{(2\pi)^2}\frac{4k_x^2}{i\omega_n+i\Omega_m-k_x^2-ak_y^2}\frac{1}{i\omega_n-k_x^2-ak_y^2}=0.
\end{equation}
The current-current correction functions in Fig1. (c) and (d) at zero temperature are
\begin{equation}
	\begin{aligned}
		\Xi_{cJ'}(i\Omega_m,0)=&-NT\sum_n\int\frac{d k^2}{(2\pi)^2}\frac{4k_x^2}{i\omega_n+i\Omega_m-k_x^2-ak_y^2}\frac{1}{(i\omega_n-k_x^2-ak_y^2)^2}\Sigma_{J'}(i\omega_n)\\
		=&\frac{2N}{\pi^2\sqrt{-a}}T\sum_n\int d k_y	\frac{\sqrt{i\omega_n+i\Omega_m+k_y^2}\mathrm{ArcTanh}(\frac{\Lambda_{UV}}{\sqrt{i\omega_n+i\Omega_m+k_y^2}})-\frac{i\omega_n+i\Omega_m+k_y^2}{\sqrt{i\omega_n+k_y^2}}\mathrm{ArcTanh}(\frac{\Lambda_{UV}}{\sqrt{i\omega_n+k_y^2}})}{\Omega_m^2} \Sigma_{J'}(i\omega_n)\\
		\approx &-\frac{J'^2N\Lambda_{U}^2\ln \Lambda}{8\pi^5(-a)}\int d\omega \frac{\omega}{\Omega_m^2}\ln(\frac{e\Lambda_{U}^2}{\tilde{J}'^2|\omega|})[\mathrm{sgn}(\omega+\Omega_m)-\mathrm{sgn}(\omega)]\\
		=&\frac{J'^2N\Lambda_{U}^2\ln \Lambda}{16\pi^5(-a)}\ln \frac{e^3\Lambda_U^4}{\tilde{J}'^4\Omega_m^2},
	\end{aligned}	
\end{equation}
and
\begin{equation}
	\begin{aligned}
		\Xi_{dJ'}(i\Omega_m,0)=& |J'|^2 N' \int\frac{d\omega_1}{2\pi}\frac{d\omega_2}{2\pi}\int\frac{d^2k_1}{(2\pi)^2}\frac{d^2k_2}{(2\pi)^2}\frac{d^2k_3}{(2\pi)^2}\frac{4k_{1x}k_{2x}}{i\omega_{1}+i\Omega_m-k_{1x}^2-ak_{1y}^2}\frac{1}{i\omega_1-k_{1x}^2-ak_{1y}^2}\frac{1}{i\omega_2+i\Omega_m-k_{2x}^2-ak_{2y}^2}\\
		&\frac{1}{i\omega_2-k_{2x}^2-ak_{2y}^2}\frac{1}{k_3^2+\tilde{J}'^2|\omega_2-\omega_1|}=0,
	\end{aligned}
\end{equation}
respectively. So the main contribution up to two-loop to electrical conductivity comes from Fig1. (c), that is
\begin{equation}
	\mathrm{Re}[\sigma(\Omega\gg T)]=\frac{\mathrm{Im}[\Xi(i\Omega_m,0)-\Xi(0,0)]_{i\Omega_m\rightarrow \Omega+i0^+}}{\Omega}=\frac{|J'|^2N\Lambda_{U}^2\ln \Lambda}{16\pi^4(-a)}\frac{1}{|\Omega|}.
\end{equation}
When $\Omega\ll T$, this leads to temperature-linear resistivity, but without a residual constant term. 

If consider the current-current correlation function arising from scattering between different patches near the van Hove points, here we make a simple assumption that the dispersion relations of these two van Hove points are denoted as $k_x^2+a k_y^2$ ($a<0$) and $b k_x^2+k_x k_y$. We also take the Fermi self-energy from Eq. (\ref{sdc-fer-2}) for simplyfy (just as demonstrated in scale analysis, the frequency dependence of the imaginary part of the Fermi self-energy due to scattering between different patch near the van Hove points remains unchanged, and only the coefficient has changed.). The relevant flow-flow correlation functions are
\begin{equation}
	\begin{aligned}
		\Xi'_b(i\Omega_m,0)-\Xi'_a(0)=&-NT\sum_n\int \frac{d^2k}{(2\pi)^2}(\frac{1}{i\omega_n+i\Omega_m-k_x^2-ak_y^2}-\frac{1}{i\omega_n-k_x^2-ak_y^2})\frac{4k_x^2}{i\omega_n-b k_x^2- k_x k_y}\\
		\approx &i\pi NT\sum_n\int \frac{d^2k}{(2\pi)^2} [\mathrm{sgn}(\omega+\Omega_m)-\mathrm{sgn}(\omega)]\delta(k_x^2+a k_y^2)\frac{4k_x^2}{i\omega_n-b k_x^2- k_x k_y}\\
		= &\frac{i\Omega_m \sqrt{-a}}{4\pi^2}[\frac{1}{ab-\sqrt{-a}}\ln \frac{e^2(ab-\sqrt{-a})^2\Lambda_U^4}{-\Omega_m^2}+\frac{1}{ab+\sqrt{-a}}\ln \frac{e^2(ab+\sqrt{-a})^2\Lambda_U^4}{-\Omega_m^2}],
	\end{aligned}
	\label{dvhp1}
\end{equation}
\begin{equation}
	\begin{aligned}
		\Xi_{cJ'}'(i\Omega_m,0)-\Xi_{bJ'}'(0,0)=&-NT\sum_n\int\frac{d k^2}{(2\pi)^2}(\frac{1}{i\omega_n+i\Omega_m-k_x^2-ak_y^2}-\frac{1}{i\omega_n-k_x^2-ak_y^2})\frac{4k_x^2}{(i\omega_n-bk_x^2-k_xk_y)^2}\Sigma_{J'}(i\omega_n)\\
		\approx &i\pi NT\sum_n\int \frac{d^2k}{(2\pi)^2} [\mathrm{sgn}(\omega+\Omega_m)-\mathrm{sgn}(\omega)]\delta(k_x^2+a k_y^2)\frac{4k_x^2}{(i\omega_n-bk_x^2-k_xk_y)^2}\Sigma_{J'}(i\omega_n)\\
		=& -i\frac{|J'|^2b\ln \Lambda}{8\pi^5(ab^2+1)}\Omega_m\ln\frac{e^4\Lambda_U^4}{\tilde{J}'^4\Omega_m^2}.
	\end{aligned}
	\label{dvhp2}	
\end{equation}
Let $\vec{k}\rightarrow -\vec{k}$, and you will also find $\Xi'_{dJ'}(i\Omega_m,0)=0$. Results in Eq. (\ref{dvhp1}) and (\ref{dvhp2}) are only valid in $ab^2+1\ne 0$. The result dc conductivity is $\mathrm{Re}[\sigma(\Omega\gg T)]=-|J'|^2b\ln (\Lambda)/8\pi^4(ab^2+1)$ and do not depend on the frequency $\Omega$.

\section{Phase diagrams}
Assuming the system exhibits superconductivity, we introduce a superconductor pairing term $g\Delta_{i j}(R,r)\psi^\dagger_i(x)\psi^\dagger_j(y)+h.c.$ [$R=(x+y)/2$ the position of the center of mass, $r=x-y$ the relative position, and $i$ and $j$ are flavor indices] into the Hamiltonian. If the fermi modes $\Psi_i$ after the Fourier-transformation primarily come from the same patch near a van Hove point, then $\Delta_{i j}(R,r)\equiv \Delta_{i j}(r)e^{i2K_1R}$ with momentum $K_1$ at this van Hove point. Interaction (\ref{sic-h}) renormalize the superconductor pair up to $\mathcal{O}(|J|^2/N)$. At the critical point of the superconductor-normal phase transition, the renormalized superconductor gap is required to be zero. In the large-$N$ limit, the linearized gap equation \cite{PhysRevLett.130.126001} is obtained as
\begin{equation}
	\Delta_{ij}(p)= \frac{|J|^2}{N}T \sum_n \int \frac{d^2k}{(2\pi)^2} G(i\omega_n,k)G(-i\omega_n,-k)\Delta_{ji}(k)D(k-p)
\end{equation}
with fermi propagator $G(i\omega_n,k)$ and boson propagator $D(k)$.  Similarly, this also applies to the case of interaction (\ref{sdc-h}), or the joint action of (\ref{sic-h}) and (\ref{sdc-h}). In the numerical computation part, we simply replace the boson propagator with $D(p)=1/(p^2+\Delta(T))$ to represent the renormalized boson propagator. Where $\Delta(T)\equiv \Delta +\tilde{\Delta}(T)$ is the square of the effective tuning parameter at finite temperature with renormalized contributions $\tilde{\Delta}(T)$.  

In the main text, we only considered the simplest case of van Hove singularties at half-filling on a square lattice. Two van Hove points are located at $(\pm \pi, 0)$ and $(0, \pm \pi)$. When electrons involved in the superconducting pairing come from the same van Hove point, the center-of-mass momentum is the reciprocal lattice vectors, and corresponds to normal superconductor. However, when electrons come from two different van Hove points, the center-of-mass momentum is $(\pm \pi, \pm \pi)$, introducing spatial modulation and leading to a density wave order. But in bilayer twisted materials, situations can arise where $2K_1$ is not a reciprocal lattice vector, leading to the appearance of the pair-density-wave order as well.

	\begin{figure}
		\centering
		\includegraphics[width=0.5\textwidth]{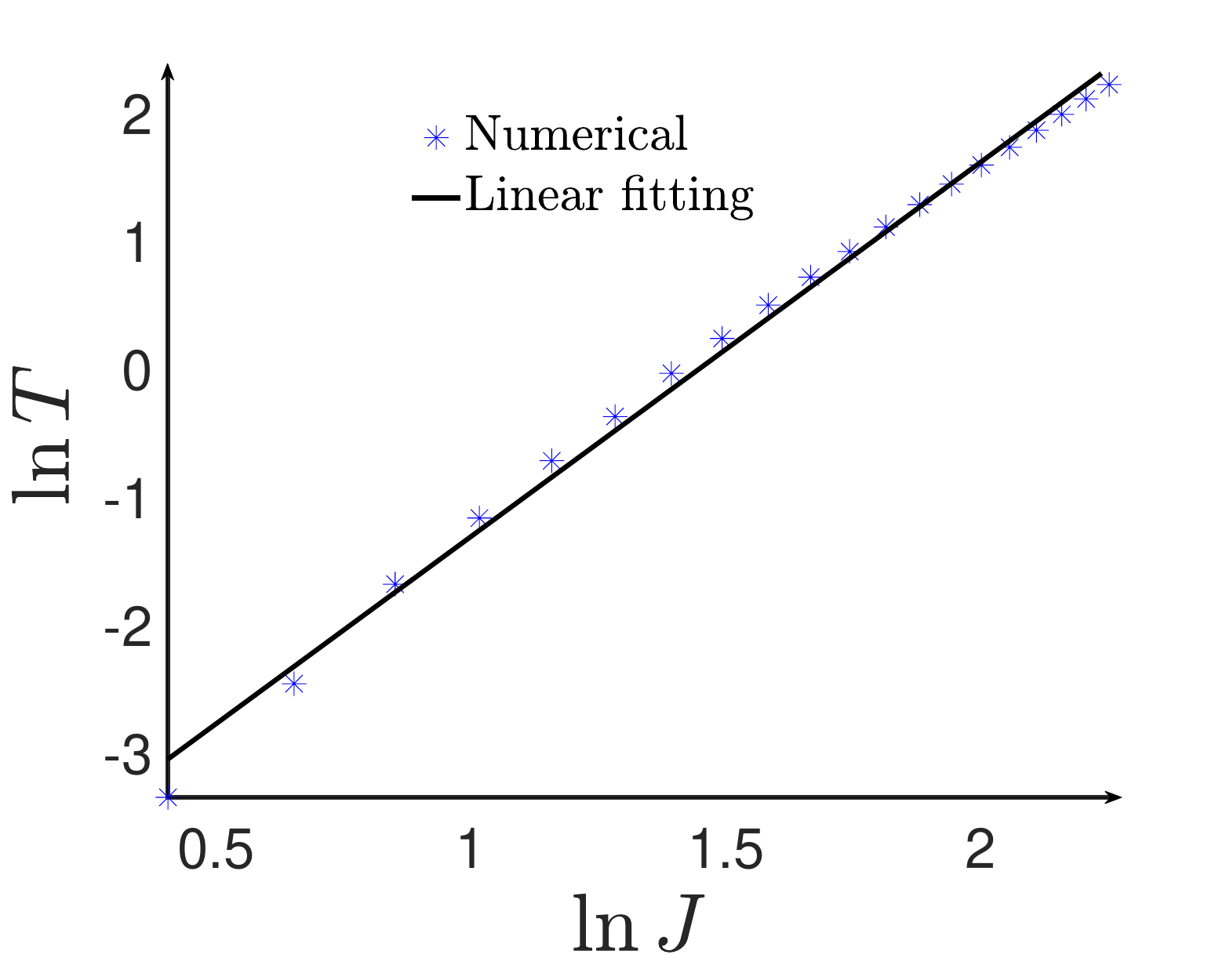}\includegraphics[width=0.5\textwidth]{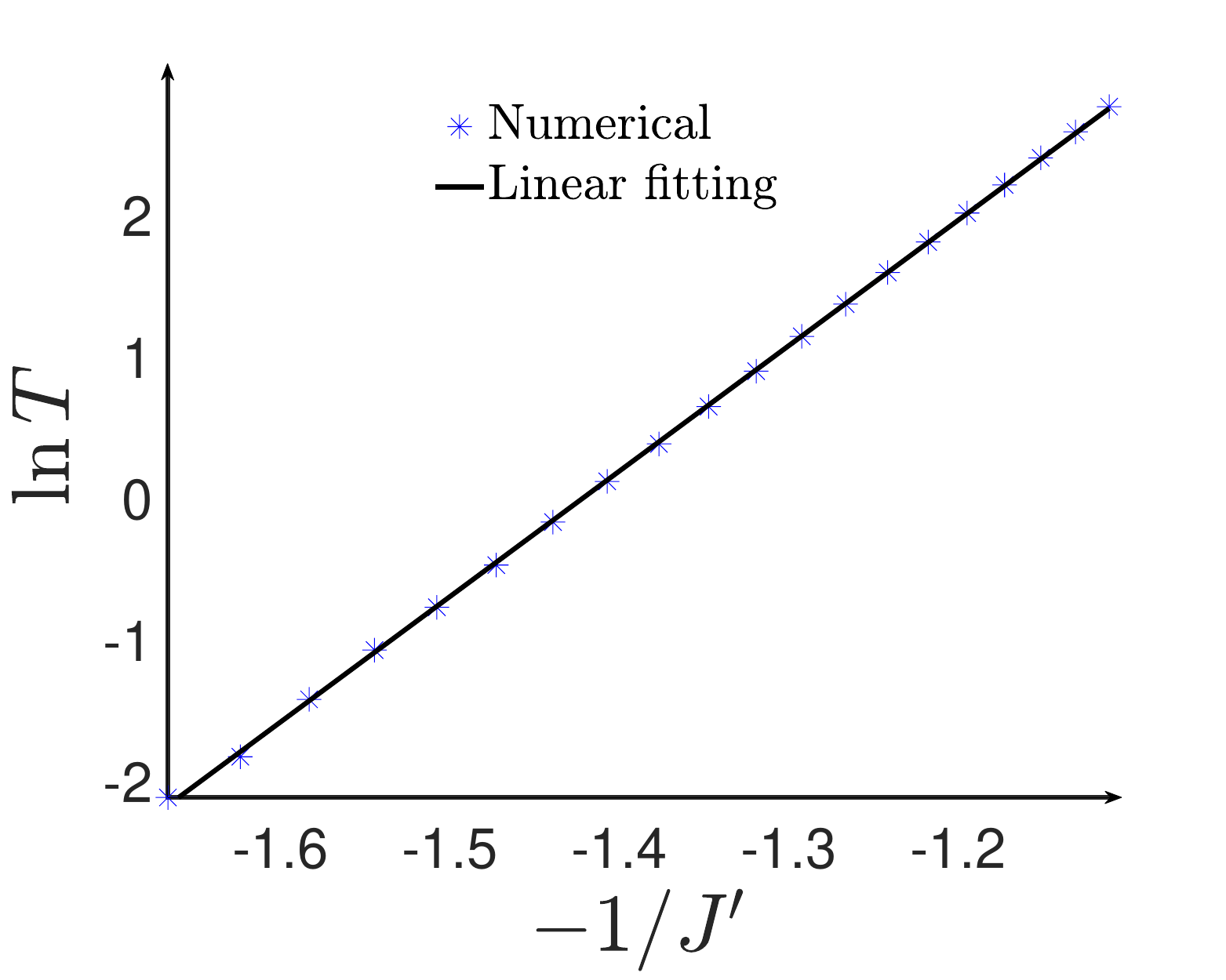}	
		\caption{Relationship of the critical temperature $T_C$ of SC with the variance of random interactions. The left diagram only consider the interaction (\ref{sic-h}) and the right one only consider interaction (\ref{sdc-h}). The parameter values are: $a=-1$, $\Lambda_U=300$, and $\Delta(T)=0.5$.} 
		\label{fig2}
	\end{figure}

	Fig. \ref{fig2}, as a supplement to Fig. 3 in the main text, provides the relationship of the critical temperature $T_C$ of superconductor with the variance of random interactions. The left graph shows a linear fit with slope $2.93$, indicating $T_C\sim (|J|/\sqrt{N})^3$, while the right graph indicates $T_C\sim e^{-\sqrt{N}/|J'|}$.

\end{document}